\documentclass[acmsmall, authorversion]{acmart}

\setcopyright{none}
\renewcommand\footnotetextcopyrightpermission[1]{}
\usepackage{algorithm}
\usepackage{algorithmic}

\usepackage{color}
\usepackage{amsmath}
\usepackage[caption=true]{subfig}
\usepackage{tabularx}

\usepackage{multirow}
\usepackage{multicol}

\usepackage{hyperref}

\newcommand{\tabincell}[2]{\begin{tabular}[t]{@{}#1@{}}#2\end{tabular}}

\AtBeginDocument{%
  \providecommand\BibTeX{{%
    \normalfont B\kern-0.5em{\scshape i\kern-0.25em b}\kern-0.8em\TeX}}}

\begin{document}

\title{The Deep Learning Compiler: A Comprehensive Survey}

\author{Mingzhen Li$^*$}
\author{Yi Liu$^*$}
\author{Xiaoyan Liu$^*$}
\author{Qingxiao Sun$^*$}
\author{Xin You$^*$}
\author{Hailong Yang$^*$$^\dagger$}
\author{Zhongzhi Luan$^*$}
\author{Lin Gan$^\mathsection$}
\author{Guangwen Yang$^\mathsection$}
\author{Depei Qian$^*$}

\affiliation{%
  \institution{Beihang University$^*$}
}

\affiliation{%
	\institution{Tsinghua University$^\mathsection$}
}

\email{{lmzhhh, yi.liu, liuxiaoyan, sunqingxiao, youxin2015, hailong.yang, zhongzhi.luan, depeiq}@buaa.edu.cn}
\email{{lingan, ygw}@tsinghua.edu.cn}

\thanks{$^\dagger$Corresponding author.}

\renewcommand{\shortauthors}{Li and Liu, et al.}

\begin{abstract}
The difficulty of deploying various deep learning (DL) models on diverse DL hardware has boosted the research and development of DL compilers in the community. Several DL compilers have been proposed from both industry and academia such as Tensorflow XLA and TVM. Similarly, the DL compilers take the DL models described in different DL frameworks as input, and then generate optimized codes for diverse DL hardware as output. However, none of the existing survey has analyzed the unique design architecture of the DL compilers comprehensively. In this paper, we perform a comprehensive survey of existing DL compilers by dissecting the commonly adopted design in details, with emphasis on the DL oriented multi-level IRs, and frontend/backend optimizations. We present detailed analysis on the design of multi-level IRs and illustrate the commonly adopted optimization techniques. Finally, several insights are highlighted as the potential research directions of DL compiler. This is the first survey paper focusing on the design architecture of DL compilers, which we hope can pave the road for future research towards DL compiler.
\end{abstract}

%

\keywords{Neural Networks, Deep Learning, Compiler, Intermediate Representation, Optimization}

\maketitle

\section{Introduction}
\label{sec:introduction}

The development of deep learning (DL) has generated profound impact on various scientific fields. It has not only demonstrated remarkable value in artificial intelligence such as natural language processing (NLP)~\cite{manning1999foundations} and computer vision (CV)~\cite{forsyth2002computer}, but also proved great success in broader applications such as e-commerce~\cite{ha2016large}, smart city~\cite{mohammadi2017semisupervised} and drug discovery~\cite{chen2018rise}. With the emergence of versatile deep learning models such as convolutional neural network (CNN)~\cite{lecun1998gradient}, recurrent neural network (RNN)~\cite{rumelhart1986learning}, long short-term memory (LSTM)~\cite{hochreiter1997long} and generative adversarial network (GAN)~\cite{goodfellow2014generative}, it is critical to ease the programming of diverse DL models in order to realize their widely adoption.

With the continuous efforts from both industry and academia, several popular DL frameworks have been proposed such as TensorFlow~\cite{abadi2016tensorflow}, PyTorch~\cite{paszke2019pytorch}, MXNet~\cite{chen2015mxnet} and CNTK~\cite{seide2016cntk}, in order to simplify the implementation of various DL models. Although there are strengths and weaknesses among the above DL frameworks depending on the tradeoffs in their designs, the interoperability becomes important to reduce the redundant engineering efforts when supporting emerging DL models across the existing DL models. To provide interoperability, ONNX~\cite{onnx} has been proposed, that defines a unified format for representing DL models to facilitate model conversion between different DL frameworks.

In the meanwhile, the unique computing characteristics such as matrix multiplication have spurred the passion of chip architects to design customized DL accelerators for higher efficiency. Internet giants (e.g., Google TPU~\cite{jouppi2017datacenter}, Hisilicon NPU~\cite{liao2019davinci}, Apple Bonic~\cite{kingsley2017inside}), processor vendors (e.g., NVIDIA Turing~\cite{turing}, Intel NNP~\cite{nnp}), service providers (e.g., Amazon Inferentia~\cite{inferentia}, Alibaba Hanguang~\cite{hanguang}), and even startups (e.g., Cambricon~\cite{7551409}, Graphcore~\cite{jia2019dissecting}) are investing tremendous workforce and capital in developing DL chips in order to boost the performance for DL models. Generally, the DL hardware can be divided into the following categories: \textit{1)} general-purpose hardware with software-hardware co-design, \textit{2)} dedicated hardware fully customized for DL models, and \textit{3)} neuromorphic hardware inspired by biological brain science. For example, the general-purpose hardware (e.g., CPU, GPU) has added special hardware components such as AVX512 vector units and tensor core to accelerate DL models. Whereas for dedicated hardware such as Google TPU, application-specific integrated circuits (e.g., matrix multiplication engine and high-bandwidth memory) have been designed to elevate the performance and energy efficiency to extreme. To the foreseeable future, the design of DL hardware would become even more diverse.

To embrace the hardware diversity, it is important to map the computation to DL hardware efficiently. On general-purpose hardware, the highly optimized linear algebra libraries such as Basic Linear Algebra Subprograms (BLAS) libraries (e.g., MKL and cuBLAS) serve as the basics for efficient computation of DL models. Take the convolution operation for example, the DL frameworks convert the convolution to matrix multiplication and then invoke the GEMM function in the BLAS libraries.
In addition, the hardware vendors have released specially optimized libraries tailored for DL computations (e.g., MKL-DNN and cuDNN), including forward and backward convolution, pooling, normalization, and activation. More advanced tools have also been developed to further speedup the DL operations. For example, TensorRT~\cite{tensorrt} supports graph optimization (e.g., layer fusion) and low-bit quantization with large collection of highly optimized GPU kernels. On dedicated DL hardware, similar libraries are also provided~\cite{7551409, jia2019dissecting}. 
However, the drawback of relying on the libraries is that they usually fall behind the rapid development of DL models, and thus fail to utilize the DL chips efficiently.

To address the drawback of DL libraries and tools, as well as alleviate the burden of optimizing the DL models on each DL hardware manually, the DL community has resorted to the domain specific compilers for rescue. Rapidly, several popular DL compilers have been proposed such as TVM~\cite{tvm}, Tensor Comprehension~\cite{tc}, Glow~\cite{glow}, nGraph~\cite{ngraph} and XLA~\cite{xla}, from both industry and academia. The DL compilers take the model definitions described in the DL frameworks as inputs, and generate efficient code implementations on various DL hardware as outputs. The transformation between model definition and specific code implementation are highly optimized targeting the model specification and hardware architecture. Specifically, they incorporate DL oriented optimizations such as layer and operator fusion, which enables highly efficient code generation. Moreover, existing DL compilers also leverage mature tool-chains from general-purpose compilers (e.g., LLVM~\cite{llvm}), which provides better portability across diverse hardware architectures. Similar to traditional compiler, DL compilers also adopt the layered design including frontend, intermediate representation (IR) and backend. However, the uniqueness of DL compiler lies in the design of multi-level IRs and DL specific optimizations.

In this paper, we provide a comprehensive survey of existing DL compilers by dissecting the compiler design into frontend, multi-level IRs and backend, with special emphasis on the IR design and optimization methods. To the best of our knowledge, this is the first paper that provides a comprehensive survey on the design of DL compiler. Specifically, this paper makes the following contributions:

\begin{itemize}
\item We dissect the commonly adopted design architecture of existing DL compilers, and provide detailed analysis of the key design components such as multi-level IRs, frontend optimizations (including node-level, block-level and dataflow-level optimizations) and backend optimizations (including hardware-specific optimization, auto-tuning and optimized kernel libraries).
\item We provide a comprehensive taxonomy of existing DL compilers from various aspects, which corresponds to the key components described in this survey. The target of this taxonomy is to provide guidelines about the selection of DL compilers for the practitioners considering their requirements, as well as to give a thorough summary of the DL compilers for researchers.
\item We have provided the quantitative performance comparison among DL compilers on CNN models, including full-fledged models and lightweight models. We have compared both end-to-end and per-layer (convolution layers since they dominate the inference time) performance to show the effectiveness of optimizations. The evaluation scripts and results are open sourced\footnote{\url{https://github.com/buaa-hipo/dlcompiler-comparison}} for reference.
\item We highlight several insights for the future development of DL compilers, including dynamic shape and pre-/post-processing, advanced auto-tuning, polyhedral model, subgraph partitioning, quantization, unified optimizations, differentiable programming and privacy protection, which we hope to boost the research in the DL compiler community.
\end{itemize}

The rest of this paper is organized as follows.
Section~\ref{sec:background} presents the background of DL compilers, including the DL frameworks, DL hardware, as well as hardware (FPGA) specific DL code generators.
Section~\ref{sec:archi_overview} describes the common design architecture of DL compilers.
Section~\ref{sec:components} discusses the key components of DL compilers, including multi-level IRs, frontend optimizations and backend optimizations.
Section~\ref{sec:taxonomy} presents a comprehensive taxonomy.
Section~\ref{sec:evaluation} provides the quantitative performance comparison.
Section~\ref{sec:conclusion} highlights the future directions for DL compiler research.

\section{Background}
\label{sec:background}

\subsection{Deep Learning Frameworks}
\label{sec:dl_frameworks}

\begin{figure}
	\centering
	\includegraphics[width=\textwidth]{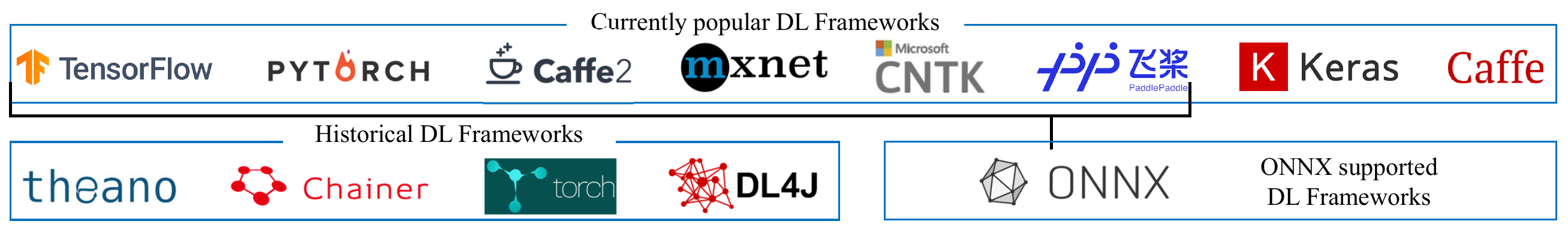}
	\caption{DL framework landscape: \textit{1)} Currently popular DL frameworks; \textit{2)} Historical DL frameworks; \textit{3)} ONNX supported frameworks.}
	\label{fig:dlframeworks}
\end{figure}

In this section, we provide an overview of popular DL frameworks. The discussion might not be exhaustive but is meant to provide a guideline fo DL practitioners. Figure~\ref{fig:dlframeworks} presents the landscape of DL frameworks including currently popular frameworks, historical frameworks and ONNX supported frameworks.

\textbf{TensorFlow -} Among all the DL frameworks, TensorFlow has the most comprehensive support for language interfaces, including C ++, Python, Java, Go, R, and Haskell. TensorFlow employs a dataflow graph of primitive operators extended with restricted control edges to represent differentiable programs~\cite{roesch2019relay}. TensorFlow Lite is designed for mobile and embedded deep learning and provides an Android neural network API. To reduce the complexity of using TensorFlow, Google adopts Keras as a frontend to the TensorFlow core. Furthermore, The eager-mode in TensorFlow applies an approach similar to PyTorch to support dynamic computation graphs better.

\textbf{Keras -} Keras~\cite{keras} is a high-level neural network library for quickly building DL models, written in pure Python. Though not a DL framework on its own, Keras provides a high-level API that integrates with TensorFlow, MXNet, Theano, and CNTK. With Keras, DL developers can build a neural network with just a few lines of code. Besides, Keras can integrate with other common DL packages, such as scikit-learn. However, Keras is not flexible enough due to over-encapsulation, which makes it too difficult to add operators or obtain low-level data information.

\textbf{PyTorch -} Facebook has rewritten the Lua-based DL framework Torch in Python and refactored all modules on \textit{Tensor} level, which leads to the release of PyTorch. As the most popular dynamic framework, PyTorch embeds primitives for constructing dynamic dataflow graphs in Python, where the control flow is executed in the Python interpreter. PyTorch 1.0 integrated the codebases of PyTorch 0.4 and Caffe2 to create a unified framework. This allows PyTorch to absorb the benefits of Caffe2 to support efficient graph execution and mobile deployment. FastAI~\cite{fastai} is an advanced API layer based on PyTorch's upper-layer encapsulation. It fully borrows Keras to ease the use of PyTorch.

\textbf{Caffe/Caffe2 -} Caffe~\cite{jia2014caffe} was designed for deep learning and image classification by UC Berkeley. Caffe has the command line, Python, and MATLAB APIs. Caffe's simplicity makes the source codes easy to extend, which is suitable for developers to analyze in-depth. Therefore, Caffe is mainly positioned in research, which has made it popular from the beginning to the present. Caffe2 is built upon the original Caffe project. Caffe2 is similar to TensorFlow in code structure, albeit with a lighter API and easier access to the intermediate results in the computation graph.

\textbf{MXNet -} MXNet supports multiple language APIs including Python, C++, R, Scala, Julia, Matlab, and JavaScript. It was intended to be scalable and was designed from the perspective to reduce data loading and I/O complexity~\cite{chen2015mxnet}. MXNet offers different paradigms: declarative programming like Caffe and Tensorflow as well as imperative like PyTorch. In December 2017, Amazon and Microsoft jointly released Gluon~\cite{gluon} based on MXNet, which is an advanced interface similar to Keras and FastAI. Gluon supports both flexible, dynamic graphs and efficient, static graphs.

\textbf{CNTK -} CNTK can be used through Python, C++ and C\# APIs, or its own scripting language (i.e., BrainScript). CNTK is designed to be easy-to-use and production-ready for large-scale data in production~\cite{hatcher2018survey}. However, CNTK does not yet support the ARM architecture, which limits its usage on mobile devices. It uses the static computation graph similar to TensorFlow and Caffe, in which a DL model is treated as a series of computational steps through a directed graph.

\textbf{PaddlePaddle -} The original design of PaddlePaddle~\cite{paddle} is similar to Caffe, where each model can be represented as a set of layers. However, PaddlePaddle v2 has adopted the concept of operators with reference to TensorFlow, which breaks layers into finer-grained operators, thereby supporting more complex DL models. And PaddlePaddle Fluid is similar to PyTorch because it provides own interpreter so as to avoid the limited performance of Python interpreter.

\textbf{ONNX -} The Open Neural Network Exchange (ONNX)~\cite{onnx} defines a scalable computation graph model, and thus computation graphs built by different DL frameworks can be easily transformed into ONNX. With ONNX, it becomes easier to convert models between DL frameworks. For example, it allows developers to build an MXNet model and then run the model using PyTorch for inference. As shown in Figure~\ref{fig:dlframeworks}, ONNX has been integrated into PyTorch, MXNet, PaddlePaddle, and so on. For several DL frameworks (e.g., TensorFlow and Keras) that are not directly supported yet, and ONNX adds converters to them.

\textbf{Historical Frameworks -} Due to the rapid evolvement in DL community, many historical DL frameworks are no longer active. For example, PyTorch has replaced Torch~\cite{torch}. As one of the oldest DL frameworks, Theano~\cite{team2016theano} is no longer under maintenance. Deeplearning4J~\cite{deeplearning4j} a distributed DL framework based on Java and Scala, however becomes inactive due to the lack of large developer community. Chainer~\cite{tokui2019chainer} was once the preferred framework for dynamic computation graphs, however replaced by MXNet, PyTorch and TensorFlow with similar features.

Previous works~\cite{bahrampour2015comparative, fonnegra2017performance, shams2017evaluation, guo2018orchestrated, nara2019performance, wei2019benchmarking} have compared the performance of DL frameworks on different applications (e.g., computer vision and image classification) and different hardware (e.g., CPU, GPU, and TPU). For detailed information about each DL framework, the readers can refer to~\cite{hatcher2018survey}. Different from them, this survey focuses on the research efforts on DL compilers which provide more general approach to execute various DL models on diverse hardware efficiently.

\subsection{Deep Learning Hardware}
\label{subsec:dl_chips}

The DL hardware can be divided into three categories based on the generality: \textit{1)} general-purpose hardware that can support DL workloads through hardware and software optimization; \textit{2)} dedicated hardware that focus on accelerating DL workloads with fully customized circuit design; \textit{3)} neuromorphic hardware that function by mimicking the human brain.

\textbf{General-purpose Hardware -} The most representative general-purpose hardware for DL models is Graphic Processing Unit (GPU), which achieves high parallelism with many-core architecture. For example, Nvidia GPUs have introduced tensor cores since the Volta architecture. Tensor cores can accelerate mixed-precision matrix multiply-and-accumulate calculations in parallel, which are widely used in DL models during both training and inference. Co-optimized with the hardware, NVIDIA also launches highly optimized DL libraries and tools such as cuDNN~\cite{chetlur2014cudnn} and TensorRT~\cite{tensorrt} to further accelerate the computation of DL models.

\textbf{Dedicated Hardware -} Dedicated hardware is fully customized for DL computation to improve performance and energy efficiency to extreme.
The rapid expansion of DL applications and algorithms has spurred many startups developing dedicated DL hardware (e.g., Graphcore GC2, Cambricon MLU270). Besides, traditional hardware companies (e.g., Intel NNP, Qualcomm Cloud AI 100) and cloud service providers (e.g., Google TPU, Amazon Inferentia, and Alibaba Hanguang) have also invested in this field.
The most well known dedicated DL hardware is Google's TPU series. A TPU includes Matrix Multiplier Unit (MXU), Unified Buffer (UB), and Activation Unit (AU), which is driven with CISC instructions by the host processor. The MXU is mainly composed of a systolic array, which is optimized for power and area efficiency in performing matrix multiplications. Compared to CPU and GPU, TPU is still programmable but uses a matrix as a primitive instead of a vector or scalar.
The Amazon Inferentia has also attracts the attention recently. This chip has four NeuroCores that are designed for tensor-level operations, and it has large on-chip cache to avoid the frequent main memory access.


\textbf{Neuromorphic Hardware -}
Neuromorphic chips use electronic technology to simulate the biological brain.
Representative products of the this kind are IBM's TrueNorth and Intel's Loihi. Neuromorphic chips (e.g., TrueNorth) have very high connectivity between their artificial neurons.
Neuromorphic chips also replicate a structure similar to the brain tissue: neurons can simultaneously store and process the data. Traditional chips distribute processors and memory in different locations, but neuromorphic chips usually have many microprocessors, each of which has a small amount of local memory.
Compared to TrueNorth, Loihi has a learning ability more similar to the brain. Loihi introduces the pulse-time-dependent synaptic plasticity model (STDP), a mechanism that regulates synaptic strength by the relative time of pre-synaptic and post-synaptic pulses.
However, neuromorphic chips are far away from Large-scale commercial production.
Despite that,  in computer science domain, neuromorphic chips can help to capture the process of rapid, life-long learning which is ignored by regular DL models, 
and in neurology domain, they are helpful to figure out how the various parts of the brain work together to create thoughts, feelings, and even consciousness.

\subsection{Hardware-specific DL Code Generator}
\label{subsec:hw_specific_dl_toolflow}

Field Programmable Gate Arrays (FPGAs) are reprogrammable integrated circuits that contain an array of programmable logic blocks. Programmers can configure them after manufacturing.
Besides the reprogrammable nature, the low-power and high-performance nature of the FPGA make it widely used in so many domains, such as communication, medical, image processing, and ASIC prototyping.
As for the domain of deep learning, the high-performance CPUs and GPUs are highly-reprogrammable but power-hungry, while the power-efficient ASICs are specialized for fixed applications. However, the FPGA can bridge the gap between CPUs/GPUs and ASICs, which causes the FPGA to be an attractive platform for deep learning.

The High-Level Synthesis (HLS) programming model enables the FPGA programmers to generate effective hardware designs conveniently using high-level languages such as C and C++. It avoids writing lots of Verilog or VHDL descriptions, which lowers the programming threshold and reduces the long design circle. Xilinx Vivado HLS and Intel FPGA SDK for OpenCL are two of the popular HLS tools targeting their own FPGAs.
However, mapping DL models to FPGAs remains a complicated work even with HLS, because that
1) DL models are usually described by the languages of DL frameworks rather than bare mental C/C++ code, and 2) DL-specific information and optimizations are hard to be leveraged.

The hardware-specific code generator targeting FPGA take the DL models or their domain-specific languages (DSLs) as the input, conduct the domain-specific (about FPGA and DL) optimizations and mappings, then generate the HLS or Verilog/VHDL and finally generate the bitstream. They can be classified into two categories according to the generated architectures of FPGA-based accelerators: the processor architecture and the streaming architecture~\cite{Venieris_2018}.

\textbf{The processor architecture} has similarities with general-purpose processors. An FPGA accelerator of this architecture usually comprises several Processing Units (PUs), which are comprised of on-chip buffers and multiple smaller Processing Engines (PEs). It usually has a virtual instruction set (ISA), and the control of hardware and the scheduling of the execution should be determined by software. What's more, the static scheduling method avoids the overheads of von Neumann execution (including instruction fetching and decoding).
A hardware template is a generic and fundamental implementation with configurable parameters. The DL code generator targeting this architecture adopt the hardware templates to generate the accelerator designs automatically. With the configurable parameters of templates, the code generator achieve the scalability and flexibility~\cite{8445088}. The scalability means that the code generator can generate designs for FPGAs ranging from high-performance to power-efficient, and the flexibility means that the code generator can generate designs for various DL models with different layer types and parameters. The number of PUs and the number of PEs per PU are template parameters of importance. Besides, the tilling size and batch size are also essential scheduling parameters about mapping the DL models to PUs and PEs. All these parameters are usually determined by the design space exploration using various strategies, such as combining the performance model and auto-tuning.
DNN Weaver~\cite{sharma2016high}, Angel-Eye~\cite{guo2017angel}, ALAMO~\cite{ma2018alamo}, FP-DNN~\cite{guan2017fp}, SysArrayAccel~\cite{wei2017automated} are typical FPGA DL code generator targeting the processor architecture.
What's more, the PUs and PEs are usually responsible for coarse-grained basic operations such as matrix-vector multiplication, matrix-matrix multiplication, pooling, and some element-wise operations. The optimizations of these basic operations are mainly guided by the tradeoff between the parallelism and data reuse, which is similar to general optimizations.

\textbf{The streaming architecture} has similarities with pipelines. An FPGA accelerator of this architecture consists of multiple different hardware blocks, and it nearly has one hardware block for each layer of an input DL model.
With the input data of a DL model, this kind of accelerators process the data through the different hardware blocks in the same sequence with layers. Additionally, with the streaming input data, all hardware blocks can be fully utilized in a pipeline manner.
However, the streaming architecture usually follows an initial assumption that the on-chip memory the computation resources on target FPGA are sufficient to accommodate the DL models, which bring barriers to deploy deep models with complicated layers.
The DL code generator targeting this architecture can solve this problem by leveraging the reconfigurability of FPGA or adopting dynamic control flow.
And the further optimization of a single block resembles that of basic operations of the processor architecture.
fpgaConvNet~\cite{venieris2016fpgaconvnet}, DeepBurning~\cite{wang2016deepburning}, Haddoc2~\cite{abdelouahab2017tactics}, and AutoCodeGen~\cite{liu2016automatic} are typical corresponding DL code generator.

For the detailed survey of specific compilation techniques that map DL models to FPGAs, the readers can refer to~\cite{Venieris_2018,8445088, guo2017survey}. Different from~\cite{Venieris_2018,8445088, guo2017survey}, this survey focuses on general DL compilation techniques that can be applied to broader DL hardware other than bounding to FPGA.

\section{Common Design Architecture of DL Compilers}
\label{sec:archi_overview}

\begin{figure*}
	\centering
	\includegraphics[scale=0.7]{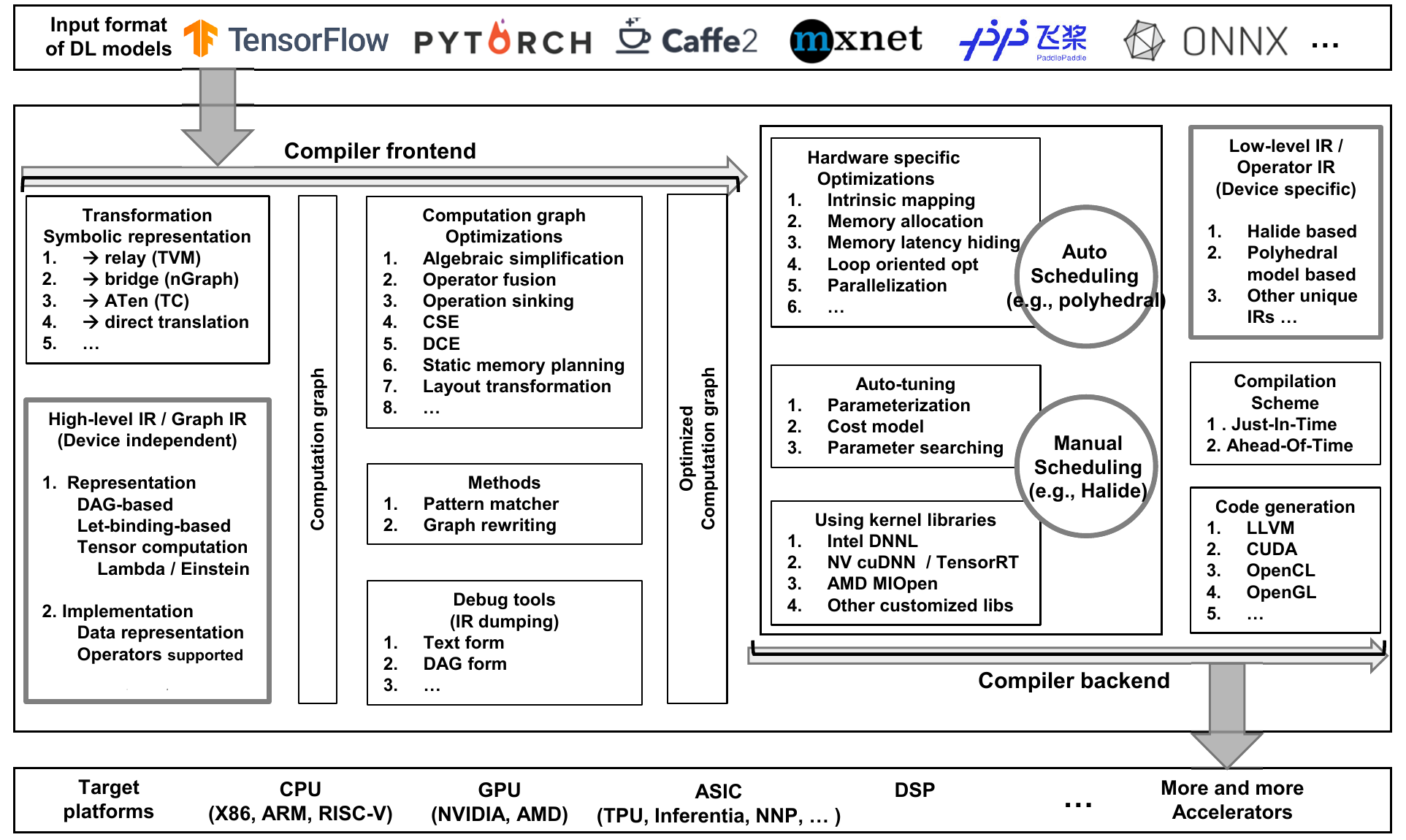}
	\caption{The overview of commonly adopted design architecture of DL compilers.}
	\label{fig:overview}
\end{figure*}

The common design architecture of a DL compiler primarily contains two parts: the compiler frontend and the compiler backend, as shown in Figure~\ref{fig:overview}. The intermediate representation (IR) is spread across both the frontend and the backend. Generally, IR is an abstraction of the program and is used for program optimizations. Specifically, the DL models are translated into multi-level IRs in DL compilers, where the high-level IR resides in the frontend, and the low-level IR resides in the backend. 
Based on the high-level IR, the compiler frontend is responsible for hardware-independent transformations and optimizations. Based on the low-level IR, the compiler backend is responsible for hardware-specific optimizations, code generation, and compilation. Note that this survey focuses on the design principles of DL compilers. For functional and experimental comparisons of DL compilers, the readers can refer to~\cite{xing2019depth,li2020deep}.

\textbf{The high-level IR}, also known as graph IR, represents the computation and the control flow and is hardware-independent.
The design challenge of high-level IR is the ability of abstraction of the computation and the control flow, which can capture and express diverse DL models. The goal of the high-level IR is to establish the control flow and the dependency between the operators and the data, as well as provide an interface for graph-level optimizations. It also contains rich semantic information for compilation as well as offers extensibility for customized operators. The detailed discussion of high-level IR is presented in Section~\ref{sec:IR_high}.

\textbf{The low-level IR} is designed for hardware-specific optimization and code generation on diverse hardware targets. Thus, the low-level IR should be fine-grained enough to reflect the hardware characteristics and represent the hardware-specific optimizations. It should also allow the use of mature third-party tool-chains in compiler backends such as Halide~\cite{ragan2013halide}, polyhedral model~\cite{polyWebsite}, and LLVM~\cite{llvm}.
The detailed discussion of low-level IR is presented in Section~\ref{sec:IR_low}.

\textbf{The frontend} takes a DL model from existing DL frameworks as input, and then transforms the model into the computation graph representation (e.g., graph IR). The frontend needs to implement various format transformations To support the diverse formats in different frameworks.
The computation graph optimizations incorporate the optimization techniques from both general-purpose compilers and the DL specific optimizations, which reduce the redundancy and improve the efficiency upon the graph IR.
Such optimizations can be classified into node-level (e.g., nop elimination and zero-dim-tensor elimination), block-level (e.g., algebraic simplification, operator fusion, and operator sinking) and dataflow-level (e.g., CSE, DCE, static memory planning, and layout transformation). After the frontend, the optimized computation graph is generated and passed to the backend. The detailed discussion of the frontend is presented in Section~\ref{subsec:frontend}. 

\textbf{The backend} transforms the high-level IR into low-level IR and performs hardware-specific optimizations.
On the one hand, it can directly transform the high-level IR to third-party tool-chains such as LLVM IR to utilize the existing infrastructures for general-purpose optimizations and code generation.
On the other hand, it can take advantage of the prior knowledge of both DL models and hardware characteristics for more efficient code generation, with customized compilation passes.
The commonly applied hardware-specific optimizations include hardware intrinsic mapping, memory allocation and fetching, memory latency hiding, parallelization as well as loop oriented optimizations.
To determine the optimal parameter setting in the large optimization space, two approaches are widely adopted in existing DL compilers such as auto-scheduling (e.g., polyhedral model) and auto-tuning (e.g., AutoTVM). The optimized low-level IR is compiled using JIT or AOT to generate codes for different hardware targets. The detailed discussion of the backend is presented in Section~\ref{sec:backend}.

\section{Key Components of DL Compilers}
\label{sec:components}

\subsection{High-level IR}
\label{sec:IR_high}

To overcome the limitation of IR adopted in traditional compilers that constrains the expression of complex computations used in DL models, existing DL compilers leverage high-level IR (as known as graph IR) with special designs for efficient code optimizations. To better understand the graph IR used in the DL compilers, we describe the representation and implementation of graph IR as follows.

\subsubsection{Representation of Graph IR}

The representation of graph IR influences the expressiveness of graph IR and also decides the way the DL compilers analyze the graph IR.

\textbf{DAG-based IR -} DAG-based IR is one of the most traditional ways for the compilers to build a computation graph, with nodes and edges organized as a directed acyclic graph (DAG). In DL compilers~\cite{tvm, ngraph, glow, tc, xla}, the nodes of a DAG represent the atomic DL operators (convolution, pooling, etc.), and the edges represent the tensors. And the graph is acyclic without loops, which differs from the data dependence graphs~\cite{dependencegraph} (DDG) of generic compilers~\cite{llvm, mlir}.
And with the help of the DAG computation graph, DL compilers can analyze the relationship and dependencies between various operators and use them to guide the optimizations. There are already plenty of optimizations on DDG, such as common sub-expression elimination (CSE) and dead code elimination (DCE). By combining the domain knowledge of DL with these algorithms, further optimizations can be applied to the DAG computation graph, which will be elaborated in Section~\ref{subsec:frontend}. DAG-based IR is convenient for programming and compiling due to its simplicity, but it has deficiencies such as semantic ambiguity caused by the missing definition of computation scope.

\textbf{Let-binding-based IR -} Let-binding is one method to solve the semantic ambiguity by offering \textit{let} expression to certain functions with restricted scope used by many high-level programming languages such as Javascript~\cite{goodman2007javascript},  F\#~\cite{petricek2012syntax}, and Scheme~\cite{abelson1998revised}. When using the \textit{let} keyword to define an expression, a \textit{let} node is generated, and then it points to the operator and variable in the expression instead of just building computational relation between variables as a DAG. In DAG-based compiler, when a process needs to get the return value of one expression, it first accesses the corresponding node and searches related nodes, also known as recursive descent technique. In contrast, the let-binding based compiler figures out all results of the variables in \textit{let} expression and builds a variable map. When a particular result is needed, the compiler looks up this map to decide the result of the expression. Among the DL compilers, the Relay IR~\cite{roesch2019relay} of TVM adopts both DAG-based IR and let-binding-based IR to obtain the benefits of both.

\textbf{Representing Tensor Computation -}
Different graph IRs have different ways to represent the computation on tensors. The operators of diverse DL frameworks are translated to graph IRs according to such specific representations. And the customized operators also need to be programmed in such representation. The representation of tensor computation can be divided into the following three categories.

\textbf{\textit{1) Function-based:}} The function-based representation just provides encapsulated operators, which is adopted by Glow, nGraph and XLA. Take High Level Optimizer (HLO, the IR of XLA) for example, it consists of a set of functions in symbolic programming, and most of them have no side-effect. The instructions are organized into three levels, including HloModule (the whole program), HloComputaion (a function), and HloInstruction (the operation). XLA uses HLO IR to represent both graph IR and operation IR so that the operation of HLO ranges from the dataflow level to the operator level.

\textbf{\textit{2) Lambda expression:}} The lambda expression, an index formula expression, describes calculation by variable binding and substitution.
Using lambda expression, programmers can define a computation quickly without implementing a new function.
TVM represents the tensor computation using the tensor expression, which is based on the lambda expression.
In TVM, computational operators in tensor expression are defined by the shape of output tensor and the lambda expression of computing rules.

\textbf{\textit{3) Einstein notation:}} The Einstein notation, also known as the summation convention, is a notation to express summation. Its programming simplicity is superior to lambda expression.
Taking TC for example, the indexes for temporary variables do not need to be defined. The IR can figure out the actual expression by the occurrence of undefined variables based on Einstein notation.  In Einstein notation, the operators need to be associative and commutative. This restriction guarantees the reduction operator can be executed by any order, making it possible for further parallelization.

\subsubsection{Implementation of Graph IR}

The implementation of graph IR in DL compilers fulfills the management of data and operation.

\textbf{Data representation -}
The data in DL compilers (e.g., inputs, weights, and intermediate data) are usually organized in the form of tensors, which are also known as multi-dimensional arrays. The DL compilers can represent tensor data directly by memory pointers, or in a more flexible way by placeholders. A placeholder contains the size for each dimension of a tensor. Alternatively, the dimension sizes of the tensor can be marked as unknown. For optimizations, the DL compilers require the data layout information. In addition, the bound of iterators should be inferred according to the placeholders.

\textbf{\textit{1) Placeholder:}} Placeholder is widely used in symbolic programming (e.g., Lisp~\cite{mccarthy1965lisp}, Tensorflow~\cite{abadi2016tensorflow}). A placeholder is simply a variable with explicit shape information (e.g., size in each dimension), and it will be populated with values at the later stage of the computation. It allows the programmers to describe the operations and build the computation graph without concerning the exact data elements, which helps separate the computation definition from the exact execution in DL compilers.
Besides, it is convenient for the programmers to change the shape of input/output and other corresponding intermediate data by using placeholders without changing the computation definition.

\textbf{\textit{2) Unknown (Dynamic) shape representation:}} The unknown dimension size is usually supported when declaring the placeholders. For instance, TVM uses \textit{Any} to represent an unknown dimension (e.g., $Tensor \langle (Any, 3), fp32 \rangle$); XLA uses \textit{None} to achieve the same purpose (e.g., $tf.placeholder$ $(``float", [None, 3])$); nGraph uses its \textit{PartialShape} class.
The unknown shape representation is necessary to support the dynamic model. However, to fully support dynamic model, the bound inference and dimension checking should be relaxed. In addition, extra mechanism should be implemented to guarantee memory validity.

\textbf{\textit{3) Data layout:}} The data layout describes how a tensor is organized in memory, and it is usually a mapping from logical indices to memory indices. The data layout usually includes the sequence of dimensions (e.g., NCHW and NHWC), tiling, padding, striding, etc.
TVM and Glow represent data layout as operator parameters and require such information for computation and optimization.
However, combining data layout information with operators rather than tensors enables intuitive implementation for certain operators and reduces the compilation overhead.
XLA represents data layout as constraints related to its backend hardware. 
Relay and MLIR are going to add data layout information into their type systems for tensors.

\textbf{\textit{4) Bound inference:}} The bound inference is applied to determine the bound of iterators when compiling DL models in DL compilers. Although the tensor representation in DL compilers is convenient to describe the inputs and outputs, it exposes special challenges for inferring the iterator bound. The bound inference is usually performed recursively or iteratively, according to the computation graph and the known placeholders.
For example, in TVM the iterators form a directed acyclic hyper-graph, where each node of the graph represents an iterator and each hyper-edge represents the relation (e.g., \textit{split}, \textit{fuse} or \textit{rebase}) among two or more iterators. Once the bound of the root iterator is determined based on the shapes of placeholders, other iterators can be inferred according to the relations recursively.

\textbf{Operators supported -}
The operators supported by DL compilers are responsible for representing the DL workloads, and they are nodes of the computation graph. The operators usually include algebraic operators (e.g., $+$, $\times$, $\exp$ and topK), neural network operators (e.g., convolution and pooling), tensor operators (e.g., reshape, resize and copy), broadcast and reduction operators (e.g., min and argmin), as well as control flow operators (e.g., conditional and loop). Here, we choose three representative operators that are frequently used across different DL compilers for illustration. In addition, we discuss the case for customized operators.

\textbf{\textit{1) Broadcast:}} The \textit{broadcast} operators can replicate the data and generate new data with compatible shape. Without \textit{broadcast} operators, the input tensor shapes are more constrained. For example, for an \textit{add} operator, the input tensors are expected to be of the same shape. Some compilers such as XLA and Relay relax such restriction by offering the \textit{broadcasting} operator. For example, XLA allows the element-wise addition on a matrix and a vector by replicating it until its shape matches the matrix.

\textbf{\textit{2) Control flow:}} Control flow is needed when representing complex and flexible models. Models such as RNN and Reinforcement learning (RL) depend on recurrent relations and data-dependent conditional execution~\cite{control-flow}, which requires control flow. Without supporting control flow in graph IR of DL compilers, these models must rely on the control flow support of the host languages (e.g., \textit{if} and \textit{while} in Python) or static unrolling, which deteriorates the computation efficiency. Relay notices that arbitrary control flow can be implemented by recursion and pattern, which has been demonstrated by functional programming~\cite{roesch2019relay}. Therefore, it provides \textit{if} operator and recursive function for implementing control flow. On the contrary, XLA represents control flow by special HLO operators such as \textit{while} and \textit{conditional}.

\textbf{\textit{3) Derivative:}} The derivative operator of an operator $Op$ takes the output gradients and the input data of $Op$ as its inputs, and then calculates the gradient of $Op$. Although some DL compilers (e.g., TVM and TC) support automatic differentiation~\cite{NIPS-AD}, they require the derivatives of all operators in high-level IR when the chain rule is applied.
TVM is working towards providing the derivative operators of both algebraic operators and neural network operators. The programmers can use these derivative operators for building the derivatives of customized operators.
On the contrary, PlaidML can generate derivative operators automatically, even for customized operators.
Notably, DL compilers unable to support derivative operators fail to provide the capability of model training.

\textbf{\textit{4) Customized operators:}} It allows programmers to define their operators for a particular purpose. Providing support for customized operators improves the extensibility of DL compilers. For example, when defining new operators in Glow, the programmers need to realize the logic and node encapsulation. In addition, extra efforts are needed, such as the lowering step, operation IR generation, and instruction generation, if necessary.
Whereas, TVM and TC require less programming efforts except describing the computation implementation. Specifically, the users of TVM only need to describe the computation and the schedule and declare the shape of input/output tensors. Moreover, the customized operators integrate Python functions through hooks, which further reduces the programmers' burden.

\subsubsection{Discussion}
Nearly all DL compilers have their unique high-level IRs. However, they share similar design philosophies, such as using DAG and let-binding to build the computation graph. In addition, they usually provide convenient ways for programmers to represent tensor computation. The data and operators designed in high-level IRs are flexible and extensible enough to support diverse DL models. More importantly, the high-level IRs are hardware-independent and thus can be applied with different hardware backend.

\subsection{Low-level IR}
\label{sec:IR_low}

\subsubsection{Implementation of Low-Level IR}
Low-level IR  describes the computation of a DL model in a more fine-grained representation than that in high-level IR, which enables the target-dependent optimizations by providing interfaces to tune the computation and memory access. In this section, we classify the common implementations of low-level IRs into three categories: Halide-based IR, polyhedral-based IR, and other unique IR. 

\textbf{Halide-based IR -}
Halide is firstly proposed to parallelize image processing, and it is proven to be extensible and efficient in DL compilers (e.g., TVM).
The fundamental philosophy of Halide is the separation of \textit{computation} and \textit{schedule}.
Rather than giving a specific scheme directly, the compilers adopting Halide try various possible \textit{schedule} and choose the best one.
The boundaries of memory reference and loop nests in Halide are restricted to bounded boxes aligned to the axes. Thus, Halide cannot express the computation with complicated patterns (e.g., non-rectangular). Fortunately, the computations in DL are quite regular to be expressed perfectly by Halide. Besides, Halide can easily parameterize these boundaries and expose them to the tuning mechanism.
The original IR of the Halide needs to be modified when applied to backend of DL compilers. For example, the input shape of Halide is infinite, whereas the DL compilers need to know the exact shape of data in order to map the operator to hardware instructions. Some compilers, such as TC, require the fixed size of data, to ensure better temporal locality for tensor data.

TVM has improved Halide IR into an independent symbolic IR by following efforts.
It removes the dependency on LLVM and refactors the structure of both the project module and the IR design of Halide, pursuing better organization as well as accessibility for graph IR and frontend language such as Python.
The re-usability is also improved, with a runtime dispatching mechanism implemented to add customized operators conveniently.
TVM simplifies the variable definition from string matching to pointer matching, guaranteeing that each variable has a single define location (static single-assignment, SSA)~\cite{cytron1991efficiently}).

\textbf{Polyhedral-based IR -}
The polyhedral model is an important technique adopted in DL compilers. It uses linear programming, affine transformations, and other mathematical methods to optimize loop-based codes with static control flow of bounds and branches.
In contrast to Halide, the boundaries of memory reference and loop nests can be polyhedrons with any shapes in the polyhedral model. Such flexibility makes polyhedral models widely used in generic compilers. However, such flexibility also prevents the integration with the tuning mechanisms. Nevertheless, due to the ability to deal with deeply nested loops, many DL compilers, such as TC and PlaidML (as the backend of nGraph), have adopted the polyhedral model as their low-level IR.
The polyhedral-based IR makes it easy to apply various polyhedral transformations (e.g., fusion, tiling, sinking, and mapping), including both device-dependent and device-independent optimizations.
There are many toolchains that are borrowed by polyhedral-based compilers, such as isl~\cite{verdoolaege2010isl}, Omega~\cite{Omega}, PIP~\cite{PIP}, Polylib~\cite{PolyLib}, and PPL~\cite{parma}.

TC has its unique design in low-level IR, which combines the Halide and polyhedral model. It uses Halide-based IR to represent the computation and adopts the polyhedral-based IR to represent the loop structures. 
TC presents detailed expressions through abstract instances and introduces specific node types. 
In brief, TC uses the \textit{domain} node to specify the ranges of index variables and uses the \textit{context} node to describe new iterative variables that are related to hardware. And it uses the \textit{band} node to determine the order of iterations.
A \textit{filter} node represents an iterator combined with a statement instance. \textit{Set} and \textit{sequence} are keywords to specify the execution types (parallel and serial execution) for \textit{filters}. Besides, TC uses \textit{extension} nodes to describe other necessary instructions for code generation, such as the memory movement. 

PlaidML uses polyhedral-based IR (called Stripe) to represent tensor operations. It creates a hierarchy of parallelizable code by extending the nesting of parallel polyhedral blocks to multiple levels. Besides, it allows nested polyhedrons to be allocated to nested memory units, providing a way to match the computation with the memory hierarchy.
In Stripe, the hardware configuration is independent of the kernel code. The \textit{tags} in Stripe (known as \textit{passes} in other compilers) do not change the kernel structure, but provide additional information about the hardware target for the optimization passes. Stripe splits the DL operators into \textit{tiles} that fit into local hardware resources.

\textbf{Other unique IR -}
There are DL compilers implementing customized low-level IRs without using Halide and polyhedral model. Upon the customized low-level IRs, they apply hardware-specific optimizations and lowers to LLVM IR.

The low-level IR in Glow is an instruction-based expression that operates on tensors referenced by addresses~\cite{glow}. There are two kinds of instruction-based functions in Glow low-level IR: \textit{declare} and \textit{program}. The first one declares the number of constant memory regions that live throughout the lifetime of the program (e.g., input, weight, bias). The second one is a list of locally allocated regions, including functions (e.g., conv and pool) and temporary variables.
Instructions can run on the global memory regions or locally allocated regions. Besides, each operand is annotated with one of the qualifiers: \textit{@in} indicates the operand reads from the buffer; \textit{@out} indicates that the operand writes to the buffer; \textit{@inout} indicates that the operand reads and writes to the buffer. These instructions and operand qualifiers help Glow determine when certain memory optimizations can be performed. 

MLIR is highly influenced by LLVM, and it is a purer compiler infrastructure than LLVM. MLIR reuses many ideas and interfaces in LLVM, and sits between the model representation and code generation.
MLIR has a flexible type system and allows multiple abstraction levels, and it introduces \textit{dialects} to represent these multiple levels of abstraction. Each \textit{dialect} consists of a set of defined immutable operations. The current \textit{dialects} of MLIR include TensorFlow IR, XLA HLO IR, experimental polyhedral IR, LLVM IR, and TensorFlow Lite. The flexible transformations between \textit{dialects} are also supported. Furthermore, MLIR can create new \textit{dialects} to connect to a new low-level compiler, which paves the way for hardware developers and compiler researchers.

The HLO IR of XLA can be considered as both high-level IR and low-level IR because HLO is fine-grained enough to represent the hardware-specific information. Besides, HLO supports hardware-specific optimizations and can be used to emit LLVM IR.

\subsubsection{Code Generation based on Low-Level IR}
\label{sec:code_gen}

The low-level IR adopted by most DL compilers can be eventually lowered to LLVM IR, and benefits from LLVM's mature optimizer and code generator.
Furthermore, LLVM can explicitly design custom instruction sets for specialized accelerators from scratch. However, traditional compilers may generate poor code when passed directly to LLVM IR. In order to avoid this situation, two approaches are applied by DL compilers to achieve hardware-dependent optimization: \textit{1)} perform target-specific loop transformation in the upper IR of LLVM (e.g., Halide-based IR and polyhedral-based IR), and \textit{2)} provide additional information about the hardware target for the optimization passes. Most DL compilers apply both approaches, but the emphasis is different. In general, the DL compilers that prefer frontend users (e.g., TC, TVM, XLA, and nGraph) might focus on \textit{1)}, whereas the DL compilers that are more inclined to backend developers (e.g., Glow, PlaidML, and MLIR) might focus on \textit{2)}.

The compilation scheme in DL compilers can be mainly classified into two categories: just-in-time (JIT) and ahead-of-time (AOT). For JIT compilers, it can generate executable codes on the fly, and they can optimize codes with better runtime knowledge.
AOT compilers generate all executable binaries first and then execute them. Thus they have a larger scope in static analysis than JIT compilation. In addition, AOT approaches can be applied with cross-compilers of embedded platforms (e.g., C-GOOD~\cite{kang2018c}) as well as enable execution on remote machines (TVM RPC) and customized accelerators.


\subsubsection{Discussion}
In DL compilers, the low-level IR is a fine-grained representation of DL models, and it reflects detailed implantation of DL models on diverse hardware. The low-level IRs include Halide-based IRs, polyhedral-based IRs, and other unique IRs. Although they differ in designs, they leverage the mature compiler tool-chains and infrastructure, to provide tailored interfaces of hardware-specific optimizations and code generation. The design of low-level IRs can also impact the design of new DL accelerators (e.g., TVM HalideIR and Inferentia, as well as XLA HLO and TPU).

\subsection{Frontend Optimizations}
\label{subsec:frontend}

After constructing the computation graph, the frontend applies graph-level optimizations. Many optimizations are easier to be identified and performed at graph level because the graph provides a global view of the computation.
These optimizations are only applied to the computation graph, rather than the implementations on backends. Thus they are hardware-independent and can be applied to various backend targets.

\begin{figure*}
	\centering
	\includegraphics[scale=0.8]{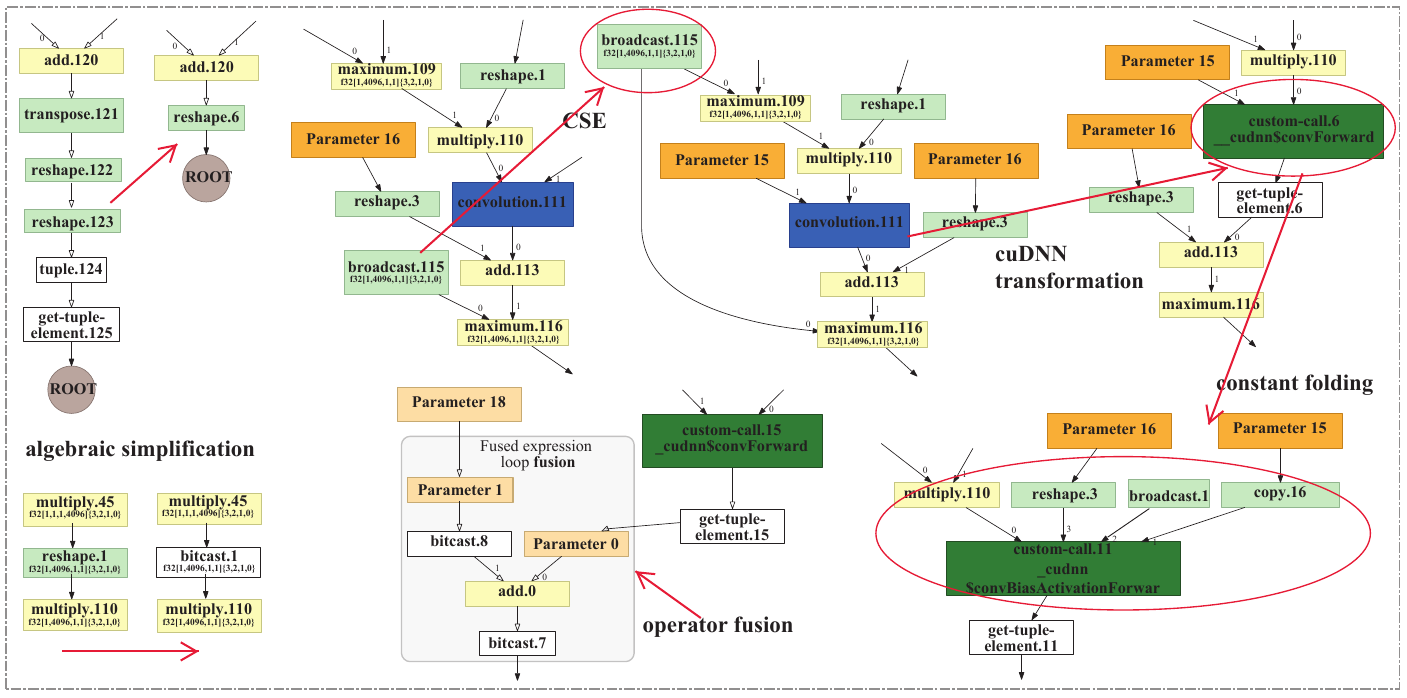}
	\caption{Example of computation graph optimizations, taken from the HLO graph of Alexnet on Volta GPU using Tensorflow XLA.}
	\label{fig:frontend-opt}
\end{figure*}

The frontend optimizations are usually defined by \textit{passes},  and can be applied by traversing the nodes of the computation graph and performing the graph transformations. The frontend provides methods to \textit{1)} capture the specific features from the computation graph and \textit{2)} rewrite the graph for optimization. Besides the pre-defined \textit{passes}, the developers can also define customized \textit{passes} in the frontend. Most DL compilers can determine the shape of both input tensors and output tensors of every operation once a DL model is imported and transformed as a computation graph. This feature allows DL compilers to perform optimizations according to the shape information. Figure~\ref{fig:frontend-opt} shows an example of computation graph optimizations with Tensorflow XLA.

In this section, we classify the frontend optimizations into three categories:
\textit{1)} node-level optimizations,
\textit{2)} block-level (peephole, local) optimizations, and
\textit{3)} dataflow-level (global) optimizations.

\subsubsection{Node-level optimizations}

The nodes of the computation graph are coarse enough to enable optimizations inside a single node.
And the node-level optimizations include node elimination that eliminates unnecessary nodes and node replacement that replaces nodes with other lower-cost nodes.

In general-purpose compilers, Nop Elimination removes the no-op instructions which occupy a small amount of space but specify no operation. In DL compilers, Nop Elimination is responsible for eliminating the operations lacking adequate inputs. For example, the \textit{sum} node with only one input tensor can be eliminated, the \textit{padding} node with zero padding width can be eliminated.

Zero-dim-tensor elimination is responsible for removing the unnecessary operations whose inputs are zero-dimension tensors. Assume that $A$ is a zero-dimension tensor, and $B$ is a constant tensor, then the sum operation node of $A$ and $B$ can be replaced with the already existing constant node $B$ without affecting the correctness. Assume that $C$ is a 3-dimension tensor, but the shape of one dimension is zero, such as \{0,2,3\}, therefore, $C$ has no element, and the argmin/argmax operation node can be eliminated.

\subsubsection{Block-level optimizations}
\textbf{Algebraic simplification -}
The algebraic simplification optimizations consist of \textit{1)}    algebraic identification, 
\textit{2)}   strength reduction, with which we can replace more expensive operators by cheaper ones; 
\textit{3)}    constant folding, with which we can replace the constant expressions by their values. 
Such optimizations consider a sequence of nodes, then take advantage of commutativity, associativity, and distributivity of different kinds of nodes to simplify the computation.

In addition to the typical operators ($+$, $\times$, etc.), the algebraic simplification can also be applied to DL specific operators (e.g., \textit{reshape}, \textit{transpose}, and \textit{pooling}). The operators can be reordered and sometimes eliminated, which reduces redundancy and improves the efficiency.
Here we illustrate the common cases where algebraic simplification can be applied: \textit{1) optimization of computation order,} in such case, the optimization finds and removes reshape/transpose operations according to specific characteristics. Taking the matrix multiplication (GEMM) for example, there are two matrices (e.g., $A$ and $B$), both matrices are transposed (to produce $A^{T}$ and $B^{T}$, respectively), then $A^{T}$ and $B^{T}$ are multiplied together. However, a more efficient way to implement GEMM is to switch the order of the arguments $A$ and $B$, multiply them together, and then transpose the output of the GEMM, which reduces two \textit{transpose} to just one; \textit{2) optimization of node combination,} in such case, the optimization combines multiple consecutive \textit{transpose} nodes into a single node, eliminates identity transpose nodes, and optimizes \textit{transpose} nodes into \textit{reshape} nodes when they actually move no data; \textit{3) optimization of ReduceMean nodes,} in such case, the optimization performs substitutions of ReduceMean with AvgPool node (e.g., in Glow), if the input of the reduce operator is 4D with the last two dimensions to be reduced.


\textbf{Operator fusion -}
Operator fusion is indispensable optimization of DL compilers. It enables better sharing of computation, eliminates intermediate allocations, facilitates further optimization by combining loop nests~\cite{roesch2019relay}, as well as reduces launch and synchronization overhead~\cite{tc}.
In TVM, the operators are classified into four categories: injective, reduction, complex-out-fusible, and opaque. When the operators are defined, their corresponding categories are determined. Targeting the above categories, TVM designs the fusion rules across operators.
In TC, fusion is performed differently based on the automatic polyhedron transformations.
However, how to identify and fuse more complicated graph patterns, such as blocks with multiple broadcast and reduce nodes, remains to be a problem. Recent works~\cite{long2018fusionstitching, long2019fusionstitching} try to tackle this problem and propose a framework to explore and optimize aggressive fusion plans. It supports not only element-wise and reduction nodes, but also other computation/memory intensive nodes with complex dependencies.

\textbf{Operator sinking -}
This optimization sinks the operations such as transposes below operations such as batch normalization, ReLU, sigmoid, and channel shuffle. By this optimization, many similar operations are moved closer to each other, creating more opportunities for algebraic simplification.

\subsubsection{Dataflow-level optimizations}
\textbf{Common sub-expression elimination (CSE) -}
An expression $E$ is a common sub-expression if the value of $E$ is previously computed, and the value of $E$ has not to be changed since previous computation~\cite{aho1986compilers}. In this case, the value of $E$ is computed once, and the already computed value of $E$ can be used to avoid recomputing in other places. The DL compilers search for common sub-expressions through the whole computation graph and replace the following common sub-expressions with the previously computed results.

\textbf{Dead code elimination (DCE) -}
A set of code is dead if its computed results or side-effects are not used. And the DCE optimization removes the dead code. The dead code is usually not caused by programmers but is caused by other graph optimizations. Thus, the DCE, as well as CSE, are applied after other graph optimizations.
Other optimizations, such as dead store elimination (DSE), which removes stores into tensors that are never going to be used, also belong to DCE.

\textbf{Static memory planning -}
Static memory planning optimizations are performed to reuse the memory buffers as much as possible. Usually, there are two approaches: in-place memory sharing and standard memory sharing. The in-place memory sharing uses the same memory for input and output for an operation, and just allocates one copy of memory before computing. Standard memory sharing reuses the memory of previous operations without overlapping. The static memory planning is done offline, which allows more complicated planning algorithms to be applied. A recent work~\cite{ahn2020ordering} firstly designs and performs memory-aware scheduling to minimize the peak activation memory footprint on edge devices, which presents new research directions of memory planning on memory-constrained devices.

\textbf{Layout transformation -}
Layout transformation tries to find the best data layouts to store tensors in the computation graph and then inserts the layout transformation nodes to the graph. Note that the actual transformation is not performed here, instead, it will be performed when evaluating the computation graph by the compiler backend. 

In fact, the performance of the same operation in different data layouts is different, and the best layouts are also different on different hardware. For example, operations in the NCHW format on GPU usually run faster, so it is efficient to transform to NCHW format on GPU (e.g., TensorFlow). Some DL compilers rely on hardware-specific libraries to achieve higher performance, and the libraries may require certain layouts. Besides, some DL accelerators prefer more complicated layouts (e.g., tile).
In addition, edge devices usually equip heterogenous computing units, and different units may require different data layouts for better utilization, thus layout transformation needs careful considerations.
Therefore, the compilers need to provide a way to perform layout transformations across various hardware.

Not only the data layouts of tensors have a nontrivial influence on the final performance, but also the transformation operations have a significant overhead. Because they also consume the memory and computation resource.

A recent work~\cite{liu2019optimizing} based on TVM targeting on CPUs alters the layout of all convolution operations to NCHW[$x$]c first in the computation graph, in which c means the split sub-dimension of channel C and $x$ indicates the split size of the sub-dimension. Then all $x$ parameters are globally explored by auto-tuning when providing hardware details, such as cache line size, vectorization unit size, and memory access pattern, during hardware-specific optimizations.

\subsubsection{Discussion}
The frontend is one of the most important components in DL compilers, which is responsible for transformation from DL models to high-level IR (e.g., computation graph) and hardware-independent optimizations based on high-level IR. Although the implementation of frontend may differ in the data representation and operator definition of high-level IR across DL compilers, the hardware-independent optimizations converge at three levels: node-level, block-level, and dataflow-level. The optimization methods at each level leverage the DL specific as well as general compilation optimization techniques, which reduce the computation redundancy as well as improve the performance of DL models at the computation graph level. 

\subsection{Backend Optimizations}
\label{sec:backend}

The backends of DL compilers have commonly included various hardware-specific optimizations, auto-tuning techniques, and optimized kernel libraries.
Hardware-specific optimizations enable efficient code generation for different hardware targets. Whereas, auto-tuning has been essential in the compiler backend to alleviate the manual efforts to derive the optimal parameter configurations. Besides, highly-optimized kernel libraries are also widely used on general-purpose processors and other customized DL accelerators.

\subsubsection{Hardware-specific Optimization}
\label{sec:backend_optimization}

\begin{figure*}
	\centering
	\includegraphics[scale=0.8]{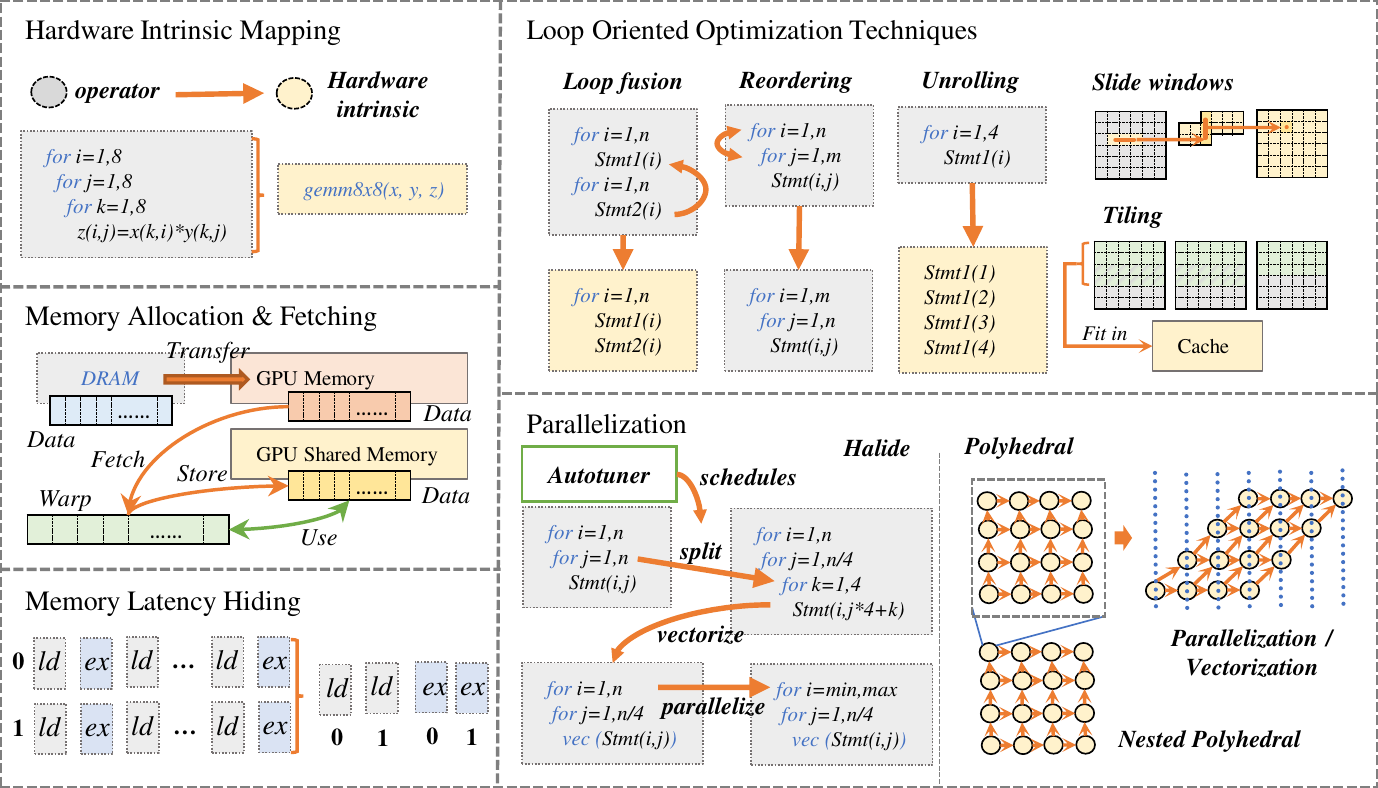}
	\caption{Overview of hardware-specific optimizations applied in DL compilers.}
	\label{fig:backend-opt}
\end{figure*}

Hardware-specific optimizations, also known as target-dependent optimizations, are applied to obtain high-performance codes targeting specific hardware. One way to apply the backend optimizations is to transform the low-level IR into LLVM IR, to utilize the LLVM infrastructure to generate optimized CPU/GPU codes. The other way is to design customized optimizations with DL domain knowledge, leveraging the target hardware more efficiently.
Since hardware-specific optimizations are tailored for particular hardware and cannot be included exhaustively in this paper, we present five widely adopted approaches in existing DL compilers.
The overview of these hardware-specific optimizations is shown in Figure~\ref{fig:backend-opt}, and the detailed descriptions are provided as follows.

\textbf{Hardware intrinsic mapping -} Hardware intrinsic mapping can transform a certain set of low-level IR instructions to kernels that have already been highly optimized on the hardware. In TVM, the hardware intrinsic mapping is realized in the method of \textit{extensible tensorization}, which can declare the behavior of hardware intrinsic and the lowering rule for intrinsic mapping. This method enables the compiler backend to apply hardware implementations as well as highly optimized handcraft micro-kernels to a specific pattern of operations, which results in a significant performance gain.
Whereas, Glow supports hardware intrinsic mapping such as \textit{quantization}. It can estimate the possible numeric range for each stage of the neural network and support profile-guided optimization to perform quantization automatically. Besides, Halide/TVM maps specific IR patterns to SIMD opcodes on each architecture to avoid the inefficiency of LLVM IR mapping when encountering vector patterns.

\textbf{Memory allocation and fetching -} Memory allocation is another challenge in code generation, especially for GPUs and customized accelerators. For example, GPU contains primarily shared memory space (lower access latency with limited memory size) and local memory space (higher access latency with large capacity). Such memory hierarchy requires efficient memory allocation and fetching techniques for improving data locality. To realize this optimization, TVM introduces the scheduling concept of \textit{memory scope}. Memory scope schedule primitives can tag a compute stage as \textit{shared} or \textit{thread-local}. For compute stages tagged as \textit{shared}, TVM generates code with shared memory allocation as well as cooperative data fetching, which inserts memory barrier at the proper code position to guarantee correctness. Besides, TC also provides similar features (known as \textit{memory promotion}) by extending PPCG~\cite{Verdoolaege2013PPCG} compiler. However, TC only supports limited pre-defined rules. Particularly, TVM enables special buffering in accelerators through \textit{memory scope} schedule primitives.

\textbf{Memory latency hiding -} Memory latency hiding is also an important technique used in the backend by reordering the execution pipeline. As most DL compilers support parallelization on CPU and GPU, memory latency hiding can be naturally achieved by hardware (e.g., warp context switching on GPU). But for TPU-like accelerators with \textit{decoupled access-execute} (DAE) architecture, the backend needs to perform scheduling and fine-grained synchronization to obtain correct and efficient codes. To achieve better performance as well as reduce programming burden, TVM introduces \textit{virtual threading} schedule primitive, which enables users to specify the data parallelism on virtualized multi-thread architecture. Then TVM lowers these virtually parallelized threads by inserting necessary memory barriers and interleaves the operations from these threads into a single instruction stream, which forms a better execution pipeline of each thread to hide the memory access latency.

\textbf{Loop oriented optimizations -} Loop oriented optimizations are also applied in the backend to generate efficient codes for target hardware. Since Halide and LLVM~\cite{llvm} (integrated with the polyhedral method) have already incorporated such optimization techniques, some DL compilers leverage Halide and LLVM in their backends. The key techniques applied in loop oriented optimizations include loop fusion, sliding windows, tiling, loop reordering, and loop unrolling.

\textbf{\textit{1) Loop fusion:}}
Loop fusion is a loop optimization technique that can fuse loops with the same boundaries for better data reuse.
For compilers such as PlaidML, TVM, TC, and XLA, such optimization is performed by the Halide schedule or polyhedral approach, while Glow applies loop fusion by its \textit{operator stacking}.

\textbf{\textit{2) Sliding windows:}}
Sliding windows is a loop optimization technique adopted by Halide. Its central concept is to compute values when needed and store them on the fly for data reuse until they are no longer required. 
As sliding windows interleaves the computation of two loops and make them serial, it is a tradeoff between parallelism and data reuse.

\textbf{\textit{3) Tiling:}}
Tiling splits loops into several tiles, and thus loops are divided into outer loops iterating through tiles and inner loops iterating inside a tile. This transformation enables better data locality inside a tile by fitting a tile into hardware caches. As the size of a tile is hardware-specific, many DL compilers determine the tiling pattern and size by auto-tuning.

\textbf{\textit{4) Loop reordering:}}
Loop reordering (also known as loop permutation) changes the order of iterations in a nested loop, which can optimize the memory access and thus increase the spatial locality. It is specific to data layout and hardware features. However, it is not safe to perform loop reordering when there are dependencies along the iteration order.

\textbf{\textit{5) Loop unrolling:}}
Loop unrolling can unroll a specific loop to a fixed number of copies of loop bodies, which allows the compilers to apply aggressive instruction-level parallelism. Usually, loop unrolling is applied in combination with loop split, which first splits the loop into two nested loops and then unrolls the inner loop completely.

\textbf{Parallelization -}
As modern processors generally support multi-threading and SIMD parallelism, the compiler backend needs to exploit parallelism to maximize hardware utilization for high performance. 
Halide uses a schedule primitive called \textit{parallel} to specify the parallelized dimension of the loop for thread-level parallelization and supports GPU parallelization by mapping loop dimensions tagged as \textit{parallel} with annotation of \textit{block} and \textit{thread}. And it replaces a loop of size $n$ with a \textit{n-wide} vector statement, which can be mapped to hardware-specific SIMD opcodes through hardware intrinsic mapping.
Stripe develops a variant of the polyhedral model called \textit{nested polyhedral model}, which introduces \textit{parallel polyhedral block} as its basic execution element of iteration. After this extension, a nested polyhedral model can detect hierarchy parallelization among levels of tiling and striding.
In addition, some DL compilers rely on handcraft libraries such as Glow or optimized math libraries provided by hardware vendors (discussed in Section~\ref{subsubsec:backend_library}). In the meanwhile, Glow offloads the vectorization to LLVM because the LLVM auto-vectorizer works well when the information of tensor dimension and loop trip count is provided. However, exploiting the parallelism entirely by compiler backend allows to apply more domain-specific knowledge of DL models, and thus leads to higher performance at the expense of more engineering efforts.



\subsubsection{Auto-tuning}
\label{subsubsec:backend_auto-tuning}

Due to the enormous search space for parameter tuning in hardware-specific optimizations, it is necessary to leverage auto-tuning to determine the optimal parameter configurations.
Among the studied DL compilers in this survey, TVM, TC, and XLA support the auto-tuning.
Generally, the auto-tuning implementation includes four key components, such as parameterization, cost model, searching technique, and acceleration.
.

\textbf{Parameterization -} \textit{1) Data and target}: The data parameter describes the specification of the data, such as input shapes. The target parameter describes hardware-specific characteristics and constraints to be considered during optimization scheduling and code generation. For example, for the GPU target, the hardware parameters such as shared memory and register size need to be specified. \textit{2) Optimization options}: The optimization options include the optimization scheduling and corresponding parameters, such as loop oriented optimizations and tile size.
In TVM, both pre-defined and user-defined scheduling, as well as parameters, are taken into consideration.
Whereas, TC and XLA  prefer to parameterize the optimizations, which have a strong correlation with performance and can be changed later at a low cost. For example, the minibatch dimension is one of the parameters that is usually mapped to grid dimensions in CUDA and can be optimized during auto-tuning.

\textbf{Cost model -} The comparison of different cost models applied in auto-tuning are as follows. 
\textit{1) Black-box model}: This model only considers the final execution time rather than the characteristics of the compilation task. It is easy to build a black-box model, but easily ends up with higher overhead and less optimal solution without the guidance of task characteristics. TC adopts this model.
\textit{2) ML-based cost model}: ML-based cost model is a statistical approach to predict performance using a machine learning method. It enables the model to update as the new configuration is explored, which helps achieve higher prediction accuracy. TVM and XLA adopt this kind of model, for example, gradient tree boosting model (GBDT) and feedforward neural network~\cite{kaufman2019learned} (FNN) respectively.
\textit{3) Pre-defined cost model}: An approach based on a pre-defined cost model expects a perfect model built on the characteristics of the compilation task and able to evaluate the overall performance of the task. 
Compared to the ML-based model, the pre-defined model generates less computation overhead when applied, but requires large engineering efforts for re-building the model on each new DL model and hardware.

\textbf{Searching technique -}
\textit{1) Initialization and searching space determination}: The initial option can either be set randomly or based on the known configurations, such as configurations given by users or historical optimal configurations.
In terms of searching space, it should be specified before auto-tuning. TVM allows developers to specify the searching space with their domain-specific knowledge and provides automatic search space extraction for each hardware target based on the computational description. In contrast, TC relies on the compilation cache and the pre-defined rules.
\textit{2) Genetic algorithm (GA)}~\cite{GA}: GA considers each tuning parameter as genes and each configuration as a candidate. The new candidate is iteratively generated by crossover, mutation, and selection according to the fitness value, which is a metaheuristic inspired by the process of natural selection. And finally, the optimal candidate is derived. The rate of crossover, mutation, and selection is used for controlling the tradeoff between exploration and exploitation. TC adopts GA in its auto-tuning technique.
\textit{3) Simulated annealing algorithm (SA)}~\cite{SA}: SA is also a metaheuristic inspired by annealing. It allows us to accept worse solutions in a decreasing probability, which can find the approximate global optimum and avoid the precise local optimum in a fixed amount of iterations. TVM adopts SA in its auto-tuning technique.
\textit{4) Reinforcement learning (RL)}: RL performs with learning to maximize reward given an environment by the tradeoff between exploration and exploitation. Chameleon~\cite{ahn2020chameleon} (built upon TVM) adopts RLRL in its auto-tuning technique.

\textbf{Acceleration -}
\textit{1) Parallelization}: One direction for accelerating auto-tuning is parallelization.
TC proposes a multi-thread, multi-GPU strategy considering that the genetic algorithm needs to evaluate all candidates in each generation. First, it enqueues candidate configurations and compiles them on multiple CPU threads. The generated code is evaluated on GPUs in parallel, and each candidate owns its fitness used by the parent choosing step. After finishing the whole evaluation, the new candidate is generated, and the new compilation job is enqueued, waiting for compiling on CPU.
Similarly, TVM supports cross-compilation and RPC, allowing users to compile on the local machine and run the programs with different auto-tuning configurations on multiple targets.
\textit{2) Configuration reuse}: Another direction for accelerating auto-tuning is to reuse the previous auto-tuning configurations. TC stores the fastest known generated code version corresponding to the given configuration by compilation cache. 
The cache is queried before each kernel optimization during the compilation, and the auto-tuning is triggered if cache miss.
Similarly, TVM produces a log file that stores the optimal configurations for all scheduling operators and queries the log file for best configurations during compilation.
It is worth mentioning that TVM performs auto-tuning for each operator in Halide IR (e.g., conv2d), and thus the optimal configurations are determined for each operator separately.

\subsubsection{Optimized Kernel Libraries}
\label{subsubsec:backend_library}
There are several highly-optimized kernel libraries widely used to accelerate DL training and inference on various hardware.
DNNL (previously MKL-DNN) from Intel, cuDNN from NVIDIA, and MIOpen from AMD are widely used libraries. Both computation-intensive primitives (e.g., convolution, GEMM, and RNN) and memory bandwidth limited primitives (e.g., batch normalization, pooling, and shuffle) are highly optimized according to the hardware features (e.g., AVX-512 ISA, tensor cores). And customizable data layouts are supported to make it easy to integrate into DL applications and avoid frequent data layout transformations. Besides, low-precision training and inference, including FP32, FP16, INT8, and non-IEEE floating-point format bfloat16~\cite{bfloat16} are also supported.
Other customized DL accelerators also maintain their specific kernel libraries~\cite{7551409, jia2019dissecting}.

Existing DL compilers, such as TVM, nGraph, and TC, can generate the function calls to these libraries during code generation.
However, if DL compilers need to leverage the existing optimized kernel libraries, they should first transform the data layouts and fusion styles into the types that are pre-defined in kernel libraries. Such transformation may break the optimal control flow. Moreover, the DL compilers treat the kernel libraries as a black box. Therefore they are unable to apply optimizations across operators (e.g., operator fusion) when invoking kernel libraries. In sum, using optimized kernel libraries achieves significant performance improvement when the computation can be satisfied by specific highly-optimized primitives, otherwise it may be constrained from further optimization and suffer from less optimal performance.

\subsubsection{Discussion}
The backend is responsible for bare-metal optimizations and code generation based on low-level IR. Although the design of backends may differ due to various low-level IRs, their optimizations can be classified into hardware-specific optimizations: auto-tuning techniques, and optimized kernel libraries.
These optimizations can be performed separately or combined, to achieve better data locality and parallelization by exploiting the hardware/software characteristics. Eventually, the high-level IR of DL models is transformed into efficient code implementation on different hardware.

\section{Taxonomy of DL Compilers}
\label{sec:taxonomy}
The DL compilers studied in this survey include TVM, nGraph, Tensor Comprehension (TC), Glow, and XLA. We select these compilers since they are well-known, well maintained, and most importantly, widely used. Thus, we can find enough papers, documents, and discussions from both industry and academia in order to study their designs and implementations in-depth.
Table~\ref{tab:taxonomy} illustrates the taxonomy of the selected DL compilers from four perspectives, including frontend, backend, IR, and optimizations, which corresponds with the key components described in this survey.

Specifically, we provide more information about the compilers to the best of our knowledge. We not only provide whether a compiler supports a specific feature, but also describe how to use this feature through its programming interface. In addition, we also describe the developing status of specific features and the reasons why specific features are not supported in particular compilers. The target of this taxonomy is to provide guidelines about the selection of DL compilers for the practitioners considering their requirements, as well as to give a thorough summary of the DL compilers for researchers.

In Table~\ref{tab:taxonomy}, we present the features of each DL compiler, including developer, programming language, ONNX/framework support, training support, and quantization support in the frontend category, and we present the compilation methods and supported devices in the backend category. These features are summarized because they strongly affect the usage of DL compilers in particular scenarios. Based on these features, practitioners or researchers can easily decide which DL compiler they would like to work upon.

Table~\ref{tab:taxonomy}, together with Figure~\ref{fig:overview} can serve as a systematic summary of this survey. Through them, readers can identify the features each compiler supports as well as the key components of each compiler. More detailed information is presented in the following sections.

\begin{table*}
	\caption{The comparison of DL compilers, including TVM, nGraph, TC, Glow, and XLA.}
	\label{tab:taxonomy}

	\renewcommand\arraystretch{1.4}
	\centering
	\scriptsize
	\begin{tabular}{p{2 pt}|p{30 pt}|p{60 pt}|p{58 pt}|p{60 pt}|p{57 pt}|p{60 pt}}
		\hline
		\hline
		&
		&
		\textbf{TVM} &
		\textbf{nGraph} &
		\textbf{TC} &
		\textbf{Glow} &
		\textbf{XLA} \\ \hline
		&
		Developer &
		Apache &
		Intel &
		Facebook &
		Facebook &
		Google \\ \hline

		\multirow{5}{*}{\rotatebox{90}{\textbf{Frontend}}} &
		\tabincell{l}{Programm-\\ing} &
		\tabincell{l}{Python/C++\\ Lambda expression} &
		\tabincell{l}{Python/C++\\ Tensor expression} &
		\tabincell{l}{Python/C++\\ Einstein notation} &
		\tabincell{l}{Python/C++\\ Layer programming} &
		\tabincell{l}{Python/C++\\ Tensorflow interface} \\ \cline{2-7}
		&
		\tabincell{l}{ONNX\\ support} &
		\tabincell{l}{\checkmark \\ tvm.relay.frontend\\ .from\_onnx (built-in)} &
		\tabincell{l}{\checkmark \\ Use ngraph-onnx\\ (Python package)} &
		$\times$ &
		\tabincell{l}{\checkmark \\ ONNXModelLoader\\ (built-in)} &
		\tabincell{l}{\checkmark \\ Use tensorflow-onnx\\ (Python package)} \\ \cline{2-7}
		&
		\tabincell{l}{Framework\\ support} &
		\tabincell{l}{tvm.relay.frontend\\.from\_* (built-in)\\ tensorflow/tflite/keras\\ pytorch/caffe2\\      mxnet/coreml/darknet} &
		\tabincell{l}{tensorflow\\ paddlepaddle\\ (Use *-bridge,\\ act as the backend) } &
		\tabincell{l}{(Define and optimize \\a TC kernel, which \\is finally called by\\ other frameworks.)\\ pytorch/other DLPack \\supported frameworks} &
		\tabincell{l}{pytorch/caffe2\\ tensorflowlite\\ (Use built-in \\ONNXIFI interface)} &
		\tabincell{l}{Use tensorflow \\interface} \\ \cline{2-7}
		&
		\tabincell{l}{Training\\ support} &
		\tabincell{l}{$\times$\\ Under developing\\ (Support derivative\\ operators now)} &
		\tabincell{l}{\checkmark \\ Only on NNP-T\\ processor} &
		\tabincell{l}{\checkmark \\ (Support auto \\differentiation)} &
		\tabincell{l}{\checkmark \\ (Limited support)} &
		\tabincell{l}{\checkmark \\ Use tensorflow\\ interface} \\ \cline{2-7}
		&
		\tabincell{l}{Quantization\\ support} &
		\tabincell{l}{\checkmark \\ int8/fp16} &
		\tabincell{l}{\checkmark \\ int8 (include training)} &
		$\times$ &
		\tabincell{l}{\checkmark \\ int8} &
		\tabincell{l}{\checkmark \\ int8/int16 (Use \\ tensorflow interface)} \\ \hline

		\multirow{2}{*}{\rotatebox{90}{\textbf{IR\ }}} &
		\tabincell{l}{High-/low- \\level IR} &
		Relay/Halide &
		nGraph IR/None &
		TC IR/Polyhedral &
		\tabincell{l}{Its own high-/low-\\ level IR} &
		\tabincell{l}{HLO (Both \\high- and low- level)} \\ \cline{2-7}
		&
		\tabincell{l}{Dynamic \\shape} &
		\tabincell{l}{\checkmark \\ (\textit{Any})} &
		\tabincell{l}{\checkmark \\ (\textit{PartialShape})} &
		\tabincell{l}{$\times$} &
		\tabincell{l}{$\times$} &
		\tabincell{l}{\checkmark \\ (\textit{None})} \\ \hline

		\multirow{4}{*}{\rotatebox{90}{\textbf{Optimization\quad}}} &
		Frontend opt &
		\multicolumn{5}{c}{\multirow{3}{*}{\tabincell{l}{Hardware independent optimizations (refer to Section~\ref{subsec:frontend})\\ Hardware specific optimizations (refer to Section~\ref{sec:backend})\\ And hybrid optimizations}}} \\  \cline{2-2}
		&
		Backend opt &
		\multicolumn{5}{c}{} \\ \cline{2-7}
		&
		Autotuning &
		\tabincell{l}{\checkmark \\ (To select the best\\ schedule   parameters)} &
		\tabincell{l}{$\times$ \\  (Call optimized kernel\\ libraries, no need)} &
		\tabincell{l}{\checkmark \\ (To reduce JIT\\ overhead)} &
		\tabincell{l}{$\times$ \\ (Additional info is \\already    provided in IR)} &
		\tabincell{l}{\checkmark \\ (On default \\ convolution and gemm )} \\ \cline{2-7}
		&
		\tabincell{l}{Kernel\\ libraries} &
		\tabincell{l}{\checkmark \\ mkl/cudnn/cublas} &
		\tabincell{l}{\checkmark \\ eigen/mkldnn/cudnn/\\Others} &
		\tabincell{l}{$\times$ } &
		\tabincell{l}{$\times$ } &
		\tabincell{l}{\checkmark \\ eigen/mkl/\\cudnn/tensorrt} \\ \hline

		\multirow{2}{*}{\rotatebox{90}{\textbf{Backend\quad}}} &
		\tabincell{l}{Compilation\\ methods} &
		\tabincell{l}{JIT\\ AOT (experimental)} &
		JIT &
		JIT &
		\tabincell{l}{JIT\\ AOT (Use built-in\\ executable bundles)} &
		\tabincell{l}{JIT\\ AOT (Generate\\ executable libraries)} \\ \cline{2-7}
		&
		Supported devices &
		\tabincell{l}{CPU/GPU/ARM\\ FPGA/Customized (\\Use VTA)} &
		\tabincell{l}{CPU/Intel GPU/NNP\\ GPU/Customized (\\Use OpenCL support\\ in PlaidML)} &
		Nvidia GPU &
		\tabincell{l}{CPU/GPU\\ Customized (\\Official docs)} &
		\tabincell{l}{CPU/GPU/TPU\\ Customized (\\Official docs)} \\ \hline

	\end{tabular}
\end{table*}

\section{Evaluation}
\label{sec:evaluation}

\subsection{Experimental Setup}

Our experiments are conducted on two GPU-equipped machines, and the hardware configuration is shown in Table~\ref{tab:setup}. We evaluate the performance of TVM (v0.6.0), nGraph (0.29.0-rc.0), TC (commit fd01443), Glow (commit 7e68188) and XLA (TensorFlow 2.2.0) on CPU and GPU. 
We select 19 neural network models in ONNX format as our datasets, which are converted from the Torchvison\footnote{\url{https://pytorch.org/docs/stable/torchvision/models.html}} model zoo and the GluonCV\footnote{\url{https://gluon-cv.mxnet.io/model_zoo/index.html}} model zoo. These models include full-fledged models: \texttt{ResNet}, \texttt{DenseNet} and \texttt{VGG} series, and lightweight models: \texttt{MobileNet} and \texttt{MNASNet} series.
To import the ONNX models, as shown in Table~\ref{tab:taxonomy}, we use the built-in \textit{tvm.relay.frontend.from\_onnx} interface of TVM, the \textit{ngraph-onnx} Python package of nGraph, the built-in \textit{ONNXModelLoader} of Glow, and the \textit{tensorflow-onnx} Python package of XLA.
Notably, TC lacks the support of ONNX, so we only evaluate it in the following per-layer performance comparison.
Each model is executed for 15 times, and we report the average execution time of the last 10 executions for each compiler, because we regard the first 5 executions as the warm-up to eliminate the overhead of JIT compilation.

\begin{table}
	\caption{The hardware configuration.}
	\label{tab:setup}

	\renewcommand\arraystretch{1.4}
	\centering
	\footnotesize
	\begin{tabular}{lll}
		\hline
		& CPU & GPU \\ \hline
		Platform a & \tabincell{l}{Broadwell E5-2680v4 *2\\ (28 physical cores, 2.4GHz)} & \tabincell{l}{Tesla V100 32GB\\ (15.7TFlops, FP32)} \\ \hline
		Platform b & \tabincell{l}{Skylake Silver 4110 *2\\ (16 physical cores, 2.1GHz)} & \tabincell{l}{Turing RTX2080Ti 11GB\\ (13.4TFlops, FP32)} \\ \hline
	\end{tabular}
\end{table}

\subsection{End-to-end Performance Comparison}

\begin{figure*}
	\centering
	\includegraphics[width=\textwidth]{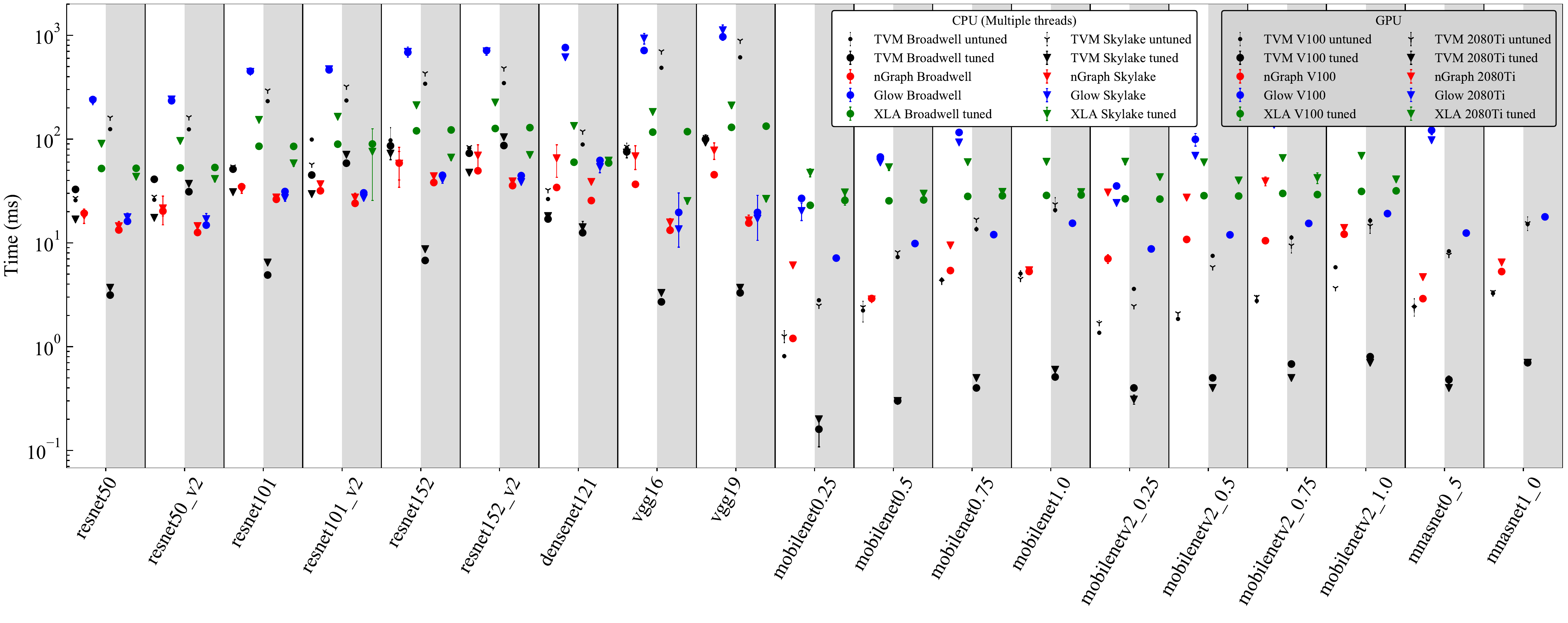}
	\caption{The performance comparison of end-to-end inference across TVM, nGraph, Glow and XLA on CPU and GPU.}
	\label{fig:e2e-perf}
\end{figure*}

As shown in Figure~\ref{fig:e2e-perf}, we compare the performance of end-to-end inference across TVM, nGraph, Glow, and XLA. We evaluate these compilers on both CPUs (Broadwell and Skylake) and GPUs (V100 and 2080Ti). Note that, we omit the comparison of TC here. Because TC is more similar to a kernel library other than fully functional DL compiler, and it requires the users to implement all layers of a model with its Einstein notion manually, which leads to heavy engineering efforts for a fair comparison. Another reason is that TC only supports running on GPU, thus we cannot obtain its performance results on CPU. However, for detailed comparisons (Figure~\ref{fig:layer-perf-mobilenet-gpu} and~\ref{fig:layer-perf-resnet-gpu}), we still implement several \texttt{ResNet} and \texttt{MobileNetV2} models in TC.
In sum, we compare and analyze the performance results from the following perspectives.

\textbf{Compatibility -}
Although nGraph and XLA claims to support ONNX , there are still compatibility problems.
\textbf{\textit{1)}} nGraph fails to run the \texttt{DenseNet121}, \texttt{VGG16/19} and \texttt{MNASNet0\_5/1\_0} models due to tensors with dynamic shapes. Alternatively, we replace the \texttt{DenseNet121}, \texttt{VGG16/19} models with the corresponding models from the ONNX model zoo\footnote{\url{https://github.com/onnx/models}}, while \texttt{MNASNet0\_5/1\_0} models are not available.
Besides, when we set PlaidML as the backend of nGraph on GPU, we fail to run all \texttt{MobileNet} models. Because PlaidML cannot handle the inconsistent definition of operators across different DL frameworks.
\textbf{\textit{2)}} XLA can run all selected models, however, the performance is quite low. Thus, we replace the selected ONNX models with the \textit{savedmodels} from the Tensorflow Hub\footnote{\url{https://tfhub.dev/}}, while the \texttt{MNASNet0\_5/1\_0} models are not available. With models from Tensorflow Hub, XLA becomes two orders of magnitude faster, and the performance of XLA becomes competitive with other compilers.

\textbf{Performance -}
From Figure~\ref{fig:e2e-perf}, we have several observations about the performance illustrated as follows.

\textbf{\textit{1)}}
\textbf{On CPU, the performance of Glow is worse than other compilers}. This is because Glow does not support thread parallelism. Thus it cannot fully utilize the multi-core CPU. Whereas TVM, nGraph, and XLA can leverage all CPU cores.

\textbf{\textit{2)}}
\textbf{XLA has the similar end-to-end inference performance for both full-fledged models (\texttt{ResNet}, \texttt{DenseNet} and \texttt{VGG} series) and lightweight models (\texttt{MobileNet} and \texttt{MNASNet} series). Besides, its inference performance on CPU and GPU is almost the same.}
It is known that XLA is embedded in the Tensorflow framework. Tensorflow contains a complicated runtime compared to TVM, nGraph, and Glow, which introduces non-trivial overhead to XLA. In addition, if we increase the batch size (set to one by default in our evaluation) and focus on the throughput of DL compilers, then the overhead of XLA can be ignored with higher throughput.

\textbf{\textit{3)}}
\textbf{In general, on CPU, TVM and nGraph achieve better performance across all models than other DL compilers}, due to the limitations of Glow and XLA described above.
TVM has comparable performance with nGraph on full-fledged models, while it is better than nGraph on lightweight models.
nGraph relies on the DNNL (previously MKL-DNN) library for acceleration. Thus, nGraph can offload the optimized subgraphs to DNNL and benefit from DNNL's fine-grained instruction-level JIT optimizations tailored for Intel CPU.

\textbf{\textit{4)}} \textbf{The tuned TVM} (tuned with 200 trials) \textbf{almost achieves the best performance on both CPU and GPU across all models, especially on lightweight models (\texttt{MobileNet}, \texttt{MNASNet} series)}. Based on our investigation, this is because the schedules of classic operators inside these models have already been well designed by TVM developers, with the default parameters provided in TVM \textit{tophub}. The default schedules and parameters can help TVM to achieve similar performance compared to other DL compilers.
In addition, the performance difference between the tuned TVM and untuned TVM is negligible on CPU but quite significant on GPU (41.26$\times$ speedup on average). This is because the GPU has more complicated thread and memory hierarchy than CPU, thus to exploit the computation power, GPU requires more fine-grained scheduling (e.g., \textit{tile}, \textit{split}, and \textit{reorder} in TVM). Therefore, it is crucial to determine the optimal scheduling parameters on GPU, where the autotuning exhibits its effectiveness.

\subsection{Per-layer Performance Comparison}
To further compare the capability of backend optimizations of DL compilers, we evaluate the per-layer (convolution layers since they dominate the inference time) performance of the \texttt{ResNet50} and \texttt{MobileNetV2$\_$1.0} on V100 GPU and Broadwell CPU (single-threaded since Glow lacks multi-threading support).

\begin{figure*}
	\centering
	\includegraphics[width=\textwidth]{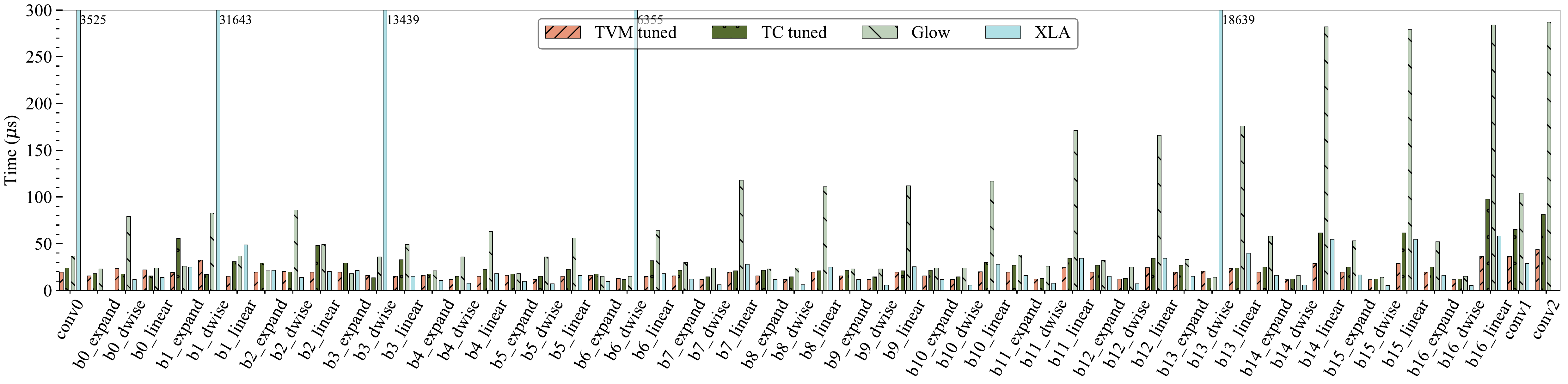}
	\caption{The performance comparison of convolution layers in \texttt{MobileNetV2$\_$1.0} across TVM, TC, Glow and XLA on V100 GPU.}
	\label{fig:layer-perf-mobilenet-gpu}
\end{figure*}

\begin{figure*}
	\centering
	\includegraphics[width=\textwidth]{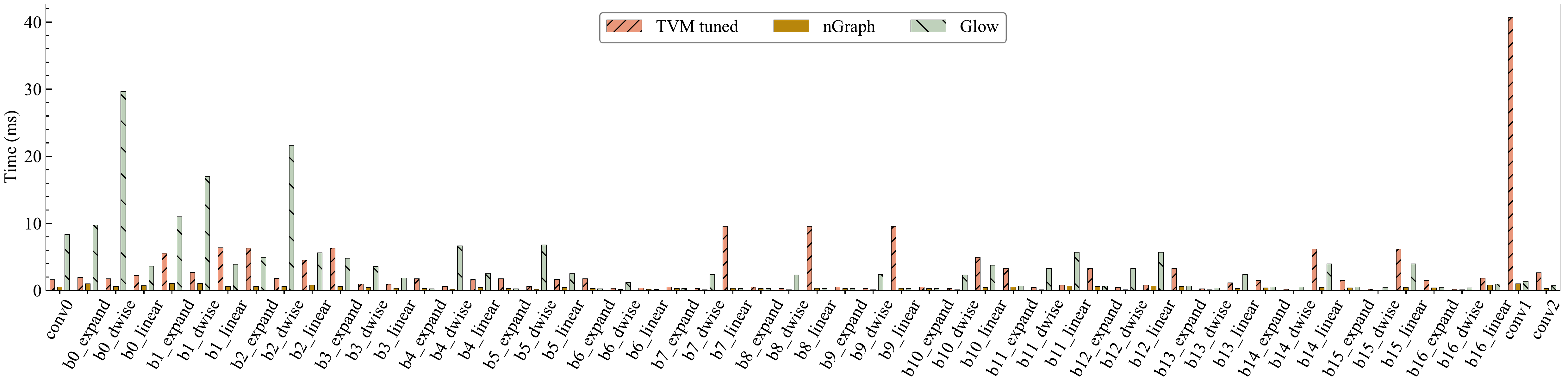}
	\caption{The performance comparison of convolution layers in \texttt{MobileNetV2$\_$1.0} across TVM, nGraph and Glow on Broadwell CPU.}
	\label{fig:layer-perf-mobilenet-cpu}
\end{figure*}

\begin{figure*}
	\centering
	\includegraphics[width=\textwidth]{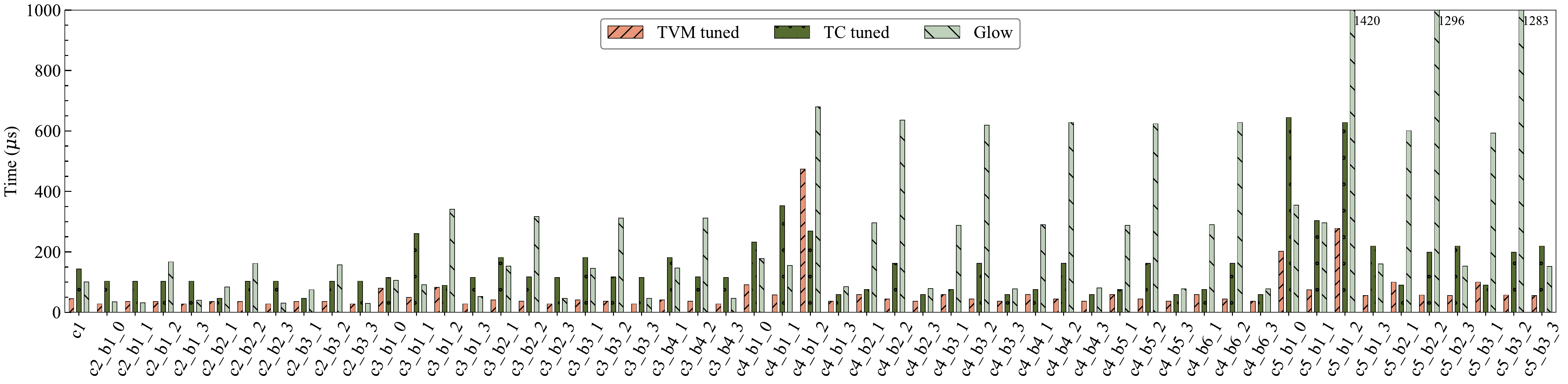}
	\caption{The performance comparison of convolution layers in \texttt{ResNet50} across TVM, TC and Glow on V100 GPU.}
	\label{fig:layer-perf-resnet-gpu}
\end{figure*}

\begin{figure*}
	\centering
	\includegraphics[width=\textwidth]{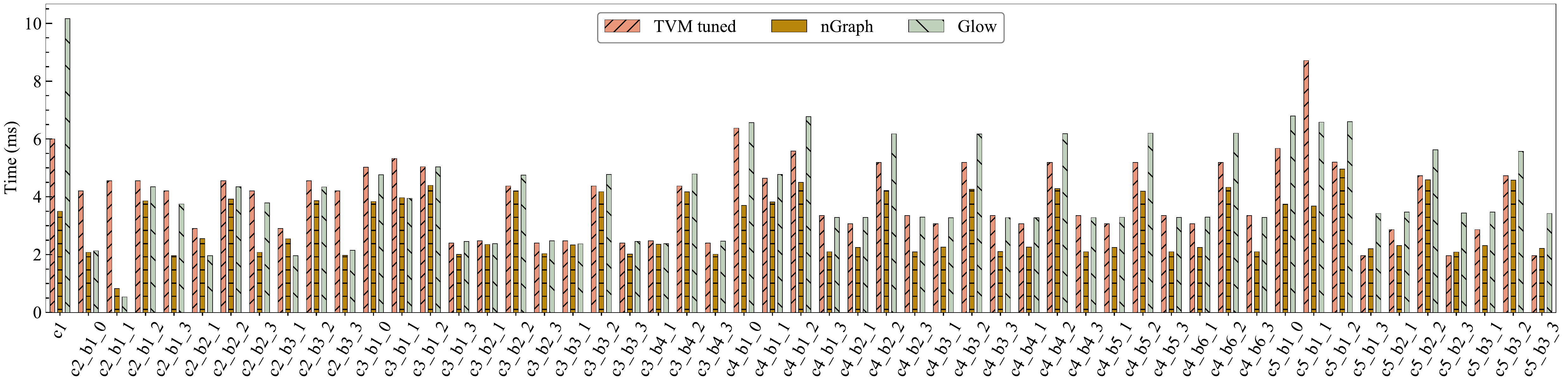}
	\caption{The performance comparison of convolution layers in \texttt{ResNet50} across TVM, nGraph and Glow on Broadwell CPU.}
	\label{fig:layer-perf-resnet-cpu}
\end{figure*}

\textbf{Methodology -}
To measure the execution time of individual layers, we adopt different methods considering the DL compilers, the hardware (CPU/GPU), and the CNN models.
Specifically, \textit{1)} On TVM, we re-use the logs of autotuning to extract the kernel shapes and the optimal schedule. Then we rebuild the individual convolution layers and use the \textit{time\_evaluator} for evaluation.
\textit{2)} We extract the execution time through the \textit{tracing} files of Glow.
\textit{3)} And we measure the execution time of hand-written kernels on TC.
\textit{4)} As for nGraph, we make use of the \textit{timeline} to measure the execution time on CPU. However, the \textit{timeline} is not supported by its PlaidML backend (which provides GPU support through OpenCL). Besides, there are no available methods to profile the command queues within OpenCL. Therefore, we leave the profiling of the per-layer performance of nGraph on GPU for future work.
\textit{4)} As for XLA, we leverage the built-in \textit{tf.profiler.experimental} method for CPU performance and the \textit{DLProf}~\cite{dlprof} toolkit from Nvidia for GPU performance.

\textbf{Performance -}
From Figure~\ref{fig:layer-perf-mobilenet-gpu},~\ref{fig:layer-perf-mobilenet-cpu}~\ref{fig:layer-perf-resnet-gpu},~\ref{fig:layer-perf-resnet-cpu}, we have several observations about the performance illustrated as follows.

\textbf{\textit{1)}}
\textbf{nGraph achieves a better performance of the convolution layers on CPU}, which benefits from the co-design of hardware (Intel CPU) and software (compiler, library, and runtime). Whereas, \textbf{TVM performs better on GPU across these compilers}. On \texttt{MobileNetV2$\_$1.0}, the performance of TVM is not stable, especially on \textit{conv1} layer. This is because the autotuning process is affected by other processes on the same machine, and thus it tends to derive the imprecise, even negative scheduling parameters.

\textbf{\textit{2)}}
TC allows users to define a tensor computation kernel (e.g., convolution) by the Einstein notion without specifying the shape of input/output tensors (e.g., kernel size). Then the kernel is autotuned and stored in its compilation cache to accelerate further autotuning and compilation. \textbf{However, in our evaluation, we find the performance of TC heavily relies on the initially compiled kernels}. Take \textit{MobileNetV2$\_$1.0} for example, if we initialize the autotuning with layer \textit{c1}, then \textit{c1} can perform well. But the following c$*\_$b$*\_*$ layers become much slower as the layers go deeper (far away from \textit{c1} layer). To derive a consistent performance, we need to tune each kernel separately.

\textbf{\textit{3)}}
\textbf{Glow falls behind other compilers to optimize the $1\times1$ convolutions} (e.g., the b$*\_$linear layers) of \texttt{MobileNetV2$\_$1.0} \textbf{as well as the depth-wise separable convolutions} (e.g., c$*\_$b$*\_$2 layers) of \texttt{ResNet50}. It takes a longer time to compute these convolutions both on GPU and CPU. We notice the convolutions are usually fused with other layers (e.g., ReLU, BatchNorm) on Glow, which could be why the lower performance compared to other compilers. Moreover, on CPU, the convolutions at the end of \texttt{MobileNetV2$\_$1.0} take a quite shorter time than convolutions at the beginning. According to the tracing log,  we notice these convolutions are accelerated by the \textit{CPUConvDKKC8} optimization~\cite{glow}, which applies tiling, layout transformation, and vectorization to convolutions with specific patterns.

\textbf{\textit{4)}}
As for XLA, it can automatically compile (\textit{\_XlaCompile}) the eligible subgraphs from Tensorflow and replace the subgraphs with the resultant binaries (\textit{\_XlaRun}). In addition, the convolution layers may be clustered with other kernels, and thus their performance is not easy to measure individually. Therefore, we have counted the clustered and the non-clustered convolutions, and the data is shown in Table~\ref{tab:xlacluster}. Note that the \texttt{MobileNetV2$\_$1.0} model in Tensorflow is a little bit different from the ONNX model for the beginning and ending layers, however, the \textit{linearbottleneck} layers are the same.
Moreover, if a convolution is to be clustered, it could be measured at most twice till the finishing of \textit{\_XlaCompile}. Therefore, there are five extreme value in Figure~\ref{fig:layer-perf-mobilenet-gpu} (corresponding with 5 clustered convolutions in \texttt{MobileNetV2$\_$1.0}).
Actually, only the clustered kernels are optimized by XLA, while the non-clustered ones are optimized by Tensorflow. Therefore, it is impossible to measure the execution time of a standalone convolution layer optimized by XLA. Consequently, we decide not to include the performance of XLA in Figure~\ref{fig:layer-perf-mobilenet-cpu} - \ref{fig:layer-perf-resnet-cpu}.

\begin{table}
	\caption{The number of the clustered and non-clustered convolutions of XLA on V100 GPU and Broadwell CPU.}
	\label{tab:xlacluster}

	\renewcommand\arraystretch{1.4}
	\centering
	\footnotesize

	\begin{tabular}{|l|l|l|l|l|}
		\hline
		\multirow{2}{*}{} & \multicolumn{2}{c|}{MobileNetV2\_1.0} & \multicolumn{2}{c|}{ResNet50} \\ \cline{2-5}
		& Clustered          & Non-clu-         & Clustered      & Non-clu-     \\ \hline
		V100              & 5                  & 47               & 0              & 53           \\ \hline
		Broadwell         & 17                 & 35               & 53             & 0            \\ \hline
	\end{tabular}
\end{table}

\subsection{Discussion}
Through the above quantitative performance comparison across DL compilers, we can in-depth analyze the coarse-grained end-to-end performance with both frontend (graph-level) and backend (operator-level) optimizations, as well as the fine-grained per-layer performance about the convolutions with backend optimizations.
However, there are still open challenges to accurately measure the effectiveness of the optimizations adopted by different DL compilers. One particular difficulty during our evaluation is that the frontend and backend optimizations are usually tightly coupled in existing DL compilers, because \textit{1)} the frontend optimizations usually affect a series of operators. Thus the optimized operators as the inputs to the backend optimizations differ across different compilers; \textit{2)} these optimizations tend to be co-designed for further exploit the performance opportunities (e.g., clustering in XLA and more advanced optimizations~\cite{long2019fusionstitching, liu2019optimizing}). Therefore, it is difficult if not impossible to evaluate and compare specific optimizations across DL compilers individually.

To tackle this problem, we have been working on building a universal benchmarking framework for existing DL compilers to measure the per-layer performance. The fundamental idea is to extract the necessary structures and parameters of the target layers (we name them as \textit{model fragments}), and rebuild the layers as acceptable inputs to a particular DL compiler, which allows the compiler to apply corresponding frontend and backend optimizations faithfully. We can then measure the performance of these optimized \textit{model fragments} to understand the effectiveness of DL compilers at layers of interests. The benchmarking framework using \textit{model fragments} is scalable to customized layers (e.g., fused layers) of interest. With such benchmarking framework available, we can derive both coarse-grained (e.g., end-to-end) and fine-grained (e.g., per-layer) performance metrics for each DL compiler, and thus compare the effectiveness of optimizations across different DL compilers at the level of interest. Currently, we have successfully experimented by extracting the target layers from the state-of-the-art CNN models, such as the \textit{bottleneck} of \texttt{ResNet50} and the \textit{linearbottleneck} of \texttt{MobileNetV2\_1.0}. Our benchmarking framework is still under rapid development, and we hope to make it available to the community soon.

\section{Conclusion and Future Directions}
\label{sec:conclusion}

In this survey, we present a thorough analysis of the existing DL compilers targeting the design principles. First, we take a deep dive into the common architecture adopted in the existing DL compilers including the multi-level IR, the frontend and the backend. We present the design philosophies and reference implementations of each component in detail, with the emphasis on the unique IRs and optimizations specific to DL compilers. We summarize the findings in this survey and highlight the future directions in DL compiler as follows:

\textbf{Dynamic shape and pre/post processing -}
Dynamic model becomes more and more popular in the field of DL, whose input shape or even model itself may change during execution. Particularly, in the area of NLP, models may accept inputs of various shapes, which is challenging for DL compilers since the shape of data is unknown until runtime. Existing DL compilers require more research efforts to support dynamic shape efficiently for emerging dynamic models.

In addition, as future DL models become more complex, their entire \textit{control flow} may inevitably include complicated pre/post-processing procedures. Currently, most DL compilers use Python as their programming language, the pre/post-processing could become a performance bottleneck when it is executed by the Python interpreter. Such potential performance bottleneck has not yet been considered by existing DL compilers.
Supporting the entire \textit{control flow} in DL compiler enables express and optimize the pre/post-processing along with DL models, which opens up new opportunities for performance acceleration in model deployment.

\textbf{Advanced auto-tuning -}
Existing auto-tuning techniques focus on the optimization of individual operators. However, the combination of the local optimal does not lead to global optimal.
For example, two adjacent operators that apply on different data layouts can be tuned together without introducing extra memory transformations in between. Besides, with the rise of edge computing, execution time is not only the optimization objective for DL compilers. New optimization targets should also be considered in the auto-tuning such as memory footprint and energy consumption.


Particularly, for the ML-based auto-tuning techniques, there are several directions worth further exploring.
First, the ML techniques can be applied in other stages of auto-tuning, other than the cost model. For example, in the stage of selecting compiler options and optimization schedules, ML techniques can be used to predict the possibility directly and develop algorithms to determine the final configurations. Second, the ML-based auto-tuning techniques can be improved based on the domain knowledge. For example, incorporating the feature engineering (selecting features to represent program)~\cite{wang2018machine} in auto-tuning techniques could be a potential direction for achieving better tuning results.



\textbf{Polyhedral model -}
It is a promising research direction to combine polyhedral model and auto-tuning techniques in the design of DL compilers for efficiency. On one hand, the auto-tuning can be applied to minimize the overhead of polyhedral JIT compilation by reusing the previous configurations. On the other hand, the polyhedral model can be used to perform auto-scheduling, which can reduce the search space of auto-tuning.

Another challenge of applying polyhedral model in DL compilers is to support the sparse tensor.  In general, the format of a sparse tensor such as CSF~\cite{smith2015tensor} expresses the loop indices with index arrays (e.g., $a[b[i]]$) that is no longer linear. Such indirect index addressing leads to non-affine subscript expressions and loop bounds, which prohibits the loop optimization of the polyhedral model~\cite{vasilache2006polyhedral,chen2012polyhedra}. Fortunately, the polyhedral community has made progress in supporting sparse tensor~\cite{venkat2014non,venkat2015loop}, and integrating the latest advancement of the polyhedral model can increase the performance opportunities for DL compilers.

\textbf{Subgraph partitioning -}
DL compilers supporting subgraph partitioning can divide the computation graph into several subgraphs, and the subgraphs can be processed in different manners. The subgraph partitioning presents more research opportunities for DL compilers.
First, it opens up the possibility to integrate graph libraries for optimization. Take nGraph and DNNL for example, DNNL is a DL library with graph optimizations leveraging vast collection of highly optimized kernels. The integration of DNNL with nGraph enables DNNL to speedup the execution of the subgraphs generated by nGraph. Secondly, it opens up the possibility of heterogeneous and parallel execution. Once the computation graph is partitioned into subgraphs, the execution of different subgraphs can be assigned to heterogeneous hardware targets at the same time. Take the edge device for example, its computation units may consist of ARM CPU, Mail GPU, DSP, and probably NPU. Generating subgraphs from the DL compilers that utilizes all computation units efficiently can deliver significant speedup of the DL tasks.



\textbf{Quantization -}
Traditional quantization strategies applied in DL frameworks are based on a set of fixed schemes and datatypes with little customization for codes running on different hardware. Whereas, supporting quantization in DL compilers can leverage optimization opportunities during compilation to derive more efficient quantization strategies. For example, Relay~\cite{roesch2019relay} provides a quantization rewriting flow that can automatically generate quantized code for various schemes.

To support quantization, there are several challenges to be solved in DL compilers. The first challenge is how to implement new quantized operators without heavy engineering efforts.
The attempt from AWS points out a possible direction that uses the concept of \textit{dialect} to implement new operators upon basic operators, so that the optimizations at graph level and operator level can be reused. The second challenge is the interaction between quantization and other optimizations during compilation. For example, determining the appropriate stage for quantization and collaborating with optimizations such as operator fusion require future research investigations.
\textbf{Unified optimizations -}
Although existing DL compilers adopt similar designs in both computation graph optimizations and hardware-specific optimizations, each compiler has its own advantages in certain aspects. There is a missing way to share the state-of-the-art optimizations, as well as support of emerging hardware targets across existing compilers. We advocate unifying the optimizations from existing DL compilers so that the best practices adopted in each DL compiler can be reused. In addition, unifying the optimizations across DL compilers can accumulate a strong force to impact the design of general-purpose and dedicated DL accelerators, and provide an environment for efficient co-design of DL compiler and hardware.

Currently, Google MLIR is a promising initiative towards such direction. It provides the infrastructure of multi-level IRs, and contains IR specification and toolkit to perform transformations across IRs at each level. It also provides flexible \textit{dialects}, so that each DL compiler can construct its customized \textit{dialects} for both high-level and low-level IRs. Through transformation across \textit{dialects}, optimizations of one DL compiler can be reused by another compiler. However, the transformation of \textit{dialects} requires further research efforts to reduce the dependency on delicate design.

\textbf{Differentiable programming -}
Differentiable programming is a programming paradigm, where the programs are differentiable thoroughly. Algorithms written in differentiable programming paradigm can be automatically differentiated, which is attractive for DL community. Many compiler projects have adopted differentiable programming, such as Myia~\cite{merrinboer2018automatic}, Flux~\cite{innes2018fashionable} and Julia~\cite{bezanson2017julia}. Unfortunately, there is little support for differential programming in existing DL compilers.


To support differential programming is quite challenging for existing DL compilers. The difficulties come from not only data structure, but also language semantic. For example, to realize the transformation from Julia to XLA HLO IR, one of the challenges~\cite{fischer2018automatic} is that the control flow is different between the imperative language used by Julia and the symbolic language used by XLA. In order to use HLO IR efficiently, the compiler also needs to provide operation abstraction for Julia in order to support the particular semantic of XLA, such as \textit{MapReduce} and \textit{broadcast}. Moreover, the semantic difference of differentiation between Julia and XLA, also requires significant changes of compiler designs.

\textbf{Privacy protection -}
In edge-cloud system, the DL models are usually split into two halves with each partial model running on the edge device and cloud service respectively, which can provide better response latency and consume less communication bandwidth. However, one of the drawbacks with the edge-cloud system is that the user privacy becomes vulnerable. The reason is that the attackers can intercept the intermediate results sent from the edge devices to cloud, and then use the intermediate results to train another model that can reveal the privacy information deviated from the original user task.

To protect privacy in edge-cloud system, existing approaches~\cite{mireshghallah2020shredder,osia2018deep,gao2019privacy} propose to add noise with special statistic properties to the intermediate results that can reduce the accuracy of the attacker task without severely deteriorating the accuracy of the user task. However, the difficulty is to determine the layer where the noise should be inserted, which is quite labor intensive to identify the optimal layer. The above difficulty presents a great opportunity for DL compilers to support privacy protection, because the compilers maintain rich information of the DL model, which can guide the noise insertion across layers automatically. 

\textbf{Training support -} In general, the model training is far less supported in current DL compilers. As shown in Table~\ref{tab:taxonomy}, nGraph only supports training on the Intel NNP-T accelerator, TC only supports the auto differentiation of a single kernel, Glow has experimental training support for limited models, the training support of TVM is under development, while XLA relies on the training support of TensorFlow. In sum, current DL compilers mainly focus on bridging the gap of deploying DL models onto diverse hardware efficiently, and thus they choose inference as their primary optimization targets. However, expanding the capability of DL compilers to support model training would open up a large body of research opportunities such as optimization of gradient operators and high-order auto differentiation.

\section*{Acknowledgements}
The authors would like to thank Jun Yang from Alibaba, Yu Xing from Xilinx, and Byung Hoon Ahn from UCSD for their valuable comments and suggestions.

%
\bibliographystyle{ACM-Reference-Format}
\bibliography{sample-base}

\end{document}